\documentclass{aa}  

\usepackage{multirow}
\usepackage{graphicx}
\usepackage{txfonts}
\usepackage{lipsum}
\usepackage{natbib}
\usepackage{booktabs}
\usepackage{amsmath}
\usepackage{comment}
\usepackage{bm}
\usepackage{tikz}
\usepackage{caption} 
\bibpunct{(}{)}{;}{a}{}{,}

\usepackage{graphicx}
\usepackage[%  
    colorlinks=true,
    pdfborder={0 0 0},
    linkcolor=blue,
    allcolors=blue
]{hyperref}
\usepackage[nameinlink]{cleveref}
\usepackage{subcaption}        
                                
\usepackage{lscape}             
\usepackage{orcidlink}    
\usepackage{placeins}           
                      
\definecolor{singleC}{HTML}{D73027}

\definecolor{10MGC}{HTML}{7FB6D8}
\definecolor{15MGC}{HTML}{3690C0}
\definecolor{16MGC}{HTML}{084594}

\newcommand{\MyCircle}{
  \begin{tikzpicture}[baseline=-0.6ex, scale=0.135]
    \draw[fill=singleC,draw=black] (0,0) circle (1);
  \end{tikzpicture}}

\newcommand{\MySquare}{
  \begin{tikzpicture}[baseline=-0.6ex, scale=0.13]
    \draw[fill=10MGC,draw=black] (-1,-1) -- (1,-1) -- (1,1) -- (-1,1) -- cycle;
  \end{tikzpicture}}

\newcommand{\MyTriangle}{
  \begin{tikzpicture}[baseline=-0.6ex, scale=0.15]
    \draw[fill=15MGC,draw=black] (90:1) -- (210:1) -- (330:1) -- cycle;
  \end{tikzpicture}}

\newcommand{\MyPentagon}{
  \begin{tikzpicture}[baseline=-0.6ex, scale=0.144]
    \draw[fill=16MGC,draw=black] (90:1)--(162:1)--(234:1)--(306:1)--(378:1)--cycle;
  \end{tikzpicture}}

\begin{document}

   \title{Magnetic Signatures in Merger Products}

   \author{N. Muntean\inst{\ref{INST:ISTA}}\orcidlink{0009-0000-1750-8618}\and L. Bugnet\inst{\ref{INST:ISTA}}\orcidlink{0000-0003-0142-4000} \and L. Buchele\inst{\ref{INST:ISTA}}\orcidlink{0000-0003-1666-4787}\and D. Dutta\inst{\ref{INST:ISTA}}\orcidlink{0009-0000-1383-8011} \and L. Einramhof\inst{\ref{INST:ISTA}}\orcidlink{0009-0002-5619-0598}\and P.\,G. Beck \inst{\ref{inst:IAC},\ref{inst:ULL}}\orcidlink{0000-0003-4745-2242}\and  L. Barrault \inst{\ref{INST:ISTA}}\orcidlink{0009-0007-7748-900X} \and B. Liagre\inst{\ref{INST:CEA}}\orcidlink{0009-0008-7869-7430}
        }

   \institute{Institute of Science and Technology Austria (ISTA), Am Campus 1, Klosterneuburg, Austria \label{INST:ISTA}
\and Instituto de Astrof\'{\i}sica de Canarias, E-38200 La Laguna, Tenerife, Spain; \label{inst:IAC}
\and Departamento de Astrof\'{\i}sica, Universidad de La Laguna, E-38206 La Laguna, Tenerife, Spain \label{inst:ULL}
 \and Universit\'e Paris Cit\'e, Universit\'e Paris-Saclay, CEA, CNRS, AIM, F-91191 Gif-sur-Yvette, France\label{INST:CEA}}

\date{Submitted June 30, 2026, Accepted September 6, 2026}

  \abstract{Asteroseismic estimates of the magnetic field strength in the radiative interior of red giant stars depend strongly on the internal stellar structure derived from models. Since red giant branch merger products have been shown to be able to possess a different core structure than single stars of the same mass, we investigate how a mass-gain history influences our estimates of an internal magnetic field strength. We construct stellar models with and without a mass-gain event after the onset of the RGB evolutionary phase with masses of $1.1\,M_\odot \le M \le 2\,M_\odot$. First, by assuming a weak magnetic field, we investigate the influence of a mass-gain event on the global sensitivity of the oscillation frequencies to the magnetic field. We find that mass-gain models can be several times more sensitive to the field than single-star models of identical total mass at masses above $1.6\,M_\odot$. Therefore, considering a mass-gain evolutionary history for merger candidates allows a significant correction to the magnetic field strength. In the presence of strong magnetic fields, we also show that the critical field strength needed to suppress mixed dipole modes is significantly lower if a mass-gain event has occurred (for stars with masses $M\gtrsim1.6\,M_\odot$). The massive end of the suppressed stars' distribution is therefore strongly favored by a merger origin. We conclude that properly constraining the stellar evolutionary history is essential when aiming to constrain internal field strength estimates from asteroseismic observations. }


   \maketitle
   \nolinenumbers

%%%%%%%%%%%%%%%%%%%%%%%%%%%%%%%%%%%%%%%%%%%%%%%%%%%%%%%%%%%%%

\section{Introduction}

The internal processes of stars spanning a large variety of masses and evolutionary stages can be measured through the observation of their oscillations \citep[e.g.][]{Aerts10} thanks to the data provided by missions like CoRoT, \textit{Kepler} and TESS \citep{Fridlund06,Borucki10,Ricker15}. The detection of mixed dipole oscillation modes in stars showing solar-like oscillations \citep[e.g.]{Beck11, Bedding11, Mosser_rot11}, and their frequency splitting due to rotation allowed for a direct constraint of the core rotation rate of stars on the red giant branch \citep[RGB, ][]{Beck12, Mosser_rot15, Vrard_rot16, Mosser_rot17, Mosser_rot18, Tayar19,Kuszlewicz23} and subgiant phase \citep{Deheuvels_rot12, Deheuvels_rot14, Deheuvels_rot16, DiMauro18, Deheuvels_rot20}. The measurements revealed that the difference between core and surface rotation of the studied stars was lower than predicted by standard hydrodynamic processes \citep[e.g.]{Beck12, Ceillier17}. This discrepancy suggests that there are additional angular momentum processes not accounted for in current models.

Besides rotation, internal magnetic fields directly affect angular momentum and chemical transport \citep{Goode92, Thompson96, Mestel88, Takahashi23}. These internal magnetic fields can be probed by seismology, as they have a perturbing effect on stellar oscillations \citep[e.g.][]{Gough90}. Thus, the inclusion of magnetic effects in purely hydrodynamic processes proves to be one of the most promising paths to fully understand the rotational measurements found for RGB stars.

The first observational evidence for possible magnetic fields in the interior of RGB stars came in the form of suppressed dipole modes \citep{Mosser_depress_12, Garcia14}. It was proposed by \cite{Fuller15} that the mode suppression stems from a trapping of the oscillation energy in the stellar core by a magnetic field. Subsequent studies \citep{Stello16a,Stello2016b} showed that suppressed dipole modes are common in red giants (RGs) that are massive enough to have possessed a convective core during the main sequence (MS). An explanation of the needed core fields to obtain such suppression effects is offered by the fossil field scenario, in which the magnetic field present during the RGB phase of a solar-like oscillator is assumed to be a remnant field created by the convective motions in the core during its MS phase \citep[e.g.][]{Cantiello16,Bugnet21}. 

Past studies on the effect of magnetism on gravity-dominated mixed modes for magnetic fields that are too weak to cause dipole mode suppression show that a present internal magnetic field introduces a frequency shift to the gravity modes, which depends on the modes' eigenfrequency \citep[e.g.][]{Gomes20, Bugnet21,Mathis21,Loi20,Loi21,Bhattacharya24}. This frequency shift causes a biased measurement of the pure dipole gravity mode period spacing ($\Delta\Pi_1$) that undershoots the true value, suggesting that magnetic fields are detectable and their strength measurable through asteroseismology \citep{Loi21, Bugnet22, Deheuvels23}. By successfully measuring magnetic frequency shifts, multiple studies reported an estimation of the weighted average of the radial magnetic field strength $\langle B_r^2\rangle^{1/2}$ (see Eq.~\ref{eq:asymptotic_mag_field_strength}) for low-mass RGs \citep[e.g.][]{Li22, Deheuvels23, Li23, Hatt24, Villate26}. Theoretical works further provided tools to better constrain the magnetic field geometry expected in stellar cores \citep[e.g.][]{Mathis23,Das24,Bhattacharya24}.

Previous works performed on magnetic detections implicitly assume stellar models that follow a standard single-star evolution in their analysis. However, a growing body of evidence suggests that many RGB stars may instead be products of merger or mass-accretion events.
In particular, the studies of \cite{Kochanek14} and \cite{deMink14} showed that merger events are very common with a predicted galactic occurrence rate of $0.2\,\mathrm{yr^{-1}}$, and that approximately $30\%$ of high-mass main-sequence stars can be classified as merger products. Synthetic population studies like \cite{Dvorakova24} further suggest that merger events leading to low- to intermediate-mass remnants should also be a common occurrence. Recent asteroseismic studies \citep[e.g.][]{Rui21, Deheuvels_mg22} report that a sizable number of \textit{Kepler} giants show asteroseismic signatures consistent with a past merger or mass-accretion event.
Indirect evidence for stellar mergers in the low- to intermediate mass range was provided by \cite{Beck2025}, who demonstrate a significant reduction of the binary rate along the RGB as orbits shorter than P$_\mathrm{orb}$\,$\lesssim$\,1000\,days do not provide sufficiently large semi-major axes to allow the giant primary to complete their H-shell burning phase without substantial star/star interaction, mass exchange or common-envelope phases.

In a seismic context, previous works have focused on the products of stellar merger or accretion events that take place during and after the MS phase of the primary \citep[e.g.][]{Rui21, Deheuvels_mg22, Henneco24, Patton25, Wu26}. The scenario of a stellar merger product produced by the merging of an RG possessing an electron-degenerate core and an MS star, both being of low- to intermediate-mass, has been studied by \cite{Rui21} and \cite{Deheuvels_mg22}. They find that if the core structure of the RG is roughly conserved in the merger remnant, unusually low values for $\Delta\Pi_1$ are expected compared to single stars with similar $\Delta\nu$. While the possible effect of stellar merger events on magnetic field detection has been partially addressed \citep{Villate26} but requires new efforts for a more in-depth characterization. A merging or mass-gain event could have two possible effects: a change to the magnetic field itself and a change in the sensitivity of stellar oscillations to the field. In this work, we focus on the second effect.

This work aims at studying the influence of a radiative interior-confined magnetic field on the properties of stellar oscillations of merger remnants. We assume an axisymmetric dipole field for low-to-intermediate mass RGB stars that subsequently experience a mass-gain/merger event. We introduce our stellar models in Sect.~\ref{sec:model_info}. In Sect.~\ref{sec:weak_fields}, we study how the inferred weighted average of the radial field strength behaves in the case of a weak field when comparing single-star and mass gain models with identical masses, mode frequencies, and magnetic frequency shifts. Finally, in Sect.~\ref{sec:strong_fields}, we study how the minimum critical field strength is influenced by an assumed mass-gain/merger event in the case of a field strong enough to cause the suppression of the gravity component of dipole mixed modes. Throughout this work, we refer to any event that increases the mass of our models as a mass-gain event, treating a stellar merger between an RGB and MS as a specific instance of a mass-gain event. 

\section{Stellar models}

\label{sec:model_info}
In this study, we consider two classes of stellar evolutionary tracks: those evolved as single stars and those that experience a mass-gain event. We simulated the mass-gain event on the early RGB, by which point the cores of the stellar models have become degenerate. The fact that our models possess a degenerate core before the onset of the mass-gain event is essential, as it results in the core structure being broadly conserved during the mass-gain event \citep{Rui21}. If the mass-gain event is considered at an earlier evolutionary stage, for example during the main-sequence, the non-degenerate model core is able to readjust to the added mass in the envelope. This leads to the mass-gain model being indistinguishable from a single star model of identical mass \citep[see Figure 4 in][]{Rui21}. All models were computed using the MESA stellar evolutionary code version 24.08.1, assuming an initial metallicity of $Z=0.02$ and no rotation. For a summary of the assumed model physics, we refer to Appendix~\ref{app:mesa_details} \footnote{The MESA project files used to calculate our models will be made available on Zenodo upon publication.} \citep{Paxton11,Paxton13,Paxton15,Paxton19,Jermyn23}. We summarize the masses of our various models in Tab.~\ref{tab:modelproperties} and provide more details below and in Appendix~\ref{app:additional_figures}.

\subsection{Single stars}
Our single-star grid spans initial masses from $1.1\,M_\odot$ to $2.0\,M_\odot$ in mass steps of $0.1\,M_\odot$. This is the typical mass range of red giant stars with detected internal magnetic fields \citep{Li22, Deheuvels23, Li23, Hatt24, Villate26, Deheuvels26}. 

\subsection{Mass gainers}

The stellar models of the mass-gain model possess initial masses of $1.0\,M_\odot,1.5\,M_\odot$, and $1.6\,M_\odot$. The chosen initial masses allow us to study both low-mass progenitors and progenitors close to the transition mass ($\sim1.5\,M_\odot$) where the behavior of the magnetic diagnostic changes as discussed in Sect.~\ref{sec:weak_fields}. We simulated a mass-gain event once the stellar model reaches the RGB, using the mass accretion functionality of MESA. To ensure that the star was fully on the RGB, the mass-gain event was initiated once the model $\nu_\mathrm{max}$ decreased below a threshold value of $\nu_\mathrm{max,thr}=250\,\mathrm{\mu Hz}$ (see Fig.~\ref{FigHRD}). The accreted material is of identical composition to the model envelope. We motivate this choice by the assumption that both stellar components that partake in a mass transfer/merger event must be formed in the same region of space, leading to minimal differences in envelope parameters like metallicity.

The $M_\mathrm{init}=1.0\,M_\odot$ models gained between $0.1$ and $1.0\,M_\odot$, while the $M_\mathrm{init}=1.5\,M_\odot$ and $M_\mathrm{init}=1.6\,M_\odot$ models gained between $0.1$ and $0.5\,M_\odot$, and $0.1$ and $0.4\,M_\odot$, respectively. In the case of all initial masses, the total gained mass was varied in steps of $0.1\,M_\odot$. During the mass-gain event, we added $10^{-5}\,M_\odot/\mathrm{yr}$ to model a rapid accretion event following \cite{Rui21}. To avoid obtaining post-gain masses above the intended threshold, we constrained the simulation timestep during the mass-gain event's duration to $250 \, \mathrm{yrs/step}$. After the cessation of the mass-gain event, we continued to evolve the resulting mass-gain products according to standard stellar evolution.
\begin{table}[t!]
\caption{Mass properties of our representative stellar models.}  
\label{tab:modelproperties}    
\centering                        
\begin{tabular}{c c c c}     
\hline\hline              
Group &Model Type & $M_\mathrm{init}\;\left[M_\odot\right]$ & $M_\mathrm{total}\;\left[M_\odot\right]$ \\         
\hline                     
   \multirow{ 2}{*}{$\left[\mathrm{a}\right]$} & single $\;\;\;\;\;\; \MyCircle$ & $1.1$ & \multirow{ 2}{*}{$1.1$} \\
   & mass-gain $\MySquare$ & $1.0$  \\ \hline
   \multirow{ 2}{*}{$\left[\mathrm{b}\right]$} & single $\;\;\;\;\;\; \MyCircle$ & $1.4$ & \multirow{ 2}{*}{$1.4$} \\
   & mass-gain $\MySquare$ & $1.0$ \\ \hline
   \multirow{ 3}{*}{$\left[\mathrm{c}\right]$} & single $\;\;\;\;\;\; \MyCircle$ & $1.6$ & \multirow{ 3}{*}{$1.6$} \\
   & mass-gain $\MySquare$ & $1.0$ \\
   & mass-gain $\MyTriangle$ & $1.5$ \\ \hline
   \multirow{4}{*}{$\left[\mathrm{d}\right]$} & single $\;\;\;\;\;\; \MyCircle$ & $2.0$ & \multirow{4}{*}{$2.0$} \\
   & mass-gain $\MySquare$ & $1.0$ \\
   & mass-gain $\MyTriangle$ & $1.5$ \\
   & mass-gain $\MyPentagon$ & $1.6$ \\
\hline                              
\end{tabular}
\tablefoot{The $M_\mathrm{init}$ and $M_\mathrm{total}$ columns represent the initial mass of the model and the model mass at the asteroseismic analysis ($\mathrm{\nu_{max}}=180\mathrm{\mu Hz}$), respectively, while the group column represents the labeling in later plots for single-star and mass-gain models of identical $M_\mathrm{total}$.}
\end{table}
For each single-star model in our grid, we computed at least one mass-gain model of equal total post-mass-gain mass. We find, that the envelope conditions (in particular the effective temperature $T_\mathrm{eff}$ and model radius $R$) are very similar for single-star and mass-gain models of identical total masses. We provide a comparison of $R$ and $T_\mathrm{eff}$ we obtained for the single-star and mass-gain models in Tab.~\ref{tab:stellar_model_params}. We extracted a stellar profile from each of the models described above at a value of $\nu_\mathrm{max} =180\,\mathrm{\mu Hz}$, a representative frequency at which magnetic signatures are commonly detected \citep[e.g.][]{Hatt24}.

\section{Influence of mass gain events on the signature of weak magnetic fields}
\label{sec:weak_fields}

We begin by investigating how a mass-gain event impacts the influence of a weak magnetic field on stellar oscillations. The magnetic fields we consider are strong enough to induce a measurable frequency shift in g-dominated dipole mixed modes near $\nu_\mathrm{max}$ of our models \citep{Bugnet21} but too weak to cause dipole mode suppression \citep[see Sect.~\ref{sec:strong_fields}, and][]{Fuller15,Stello2016b}. %For each of our stellar models, we aim to study the impact of an assumed mass-gain history on the sensitivity of the oscillation modes to the magnetic field. 
In the following section, we briefly present the definition and meaning of the magnetic seismic parameters used in this part of the study.

\subsection{Frequency perturbation due to weak magnetic effects}
\label{sec:magnetic_theory}

\begin{figure*}
    \sidecaption
    \includegraphics[width=12cm]{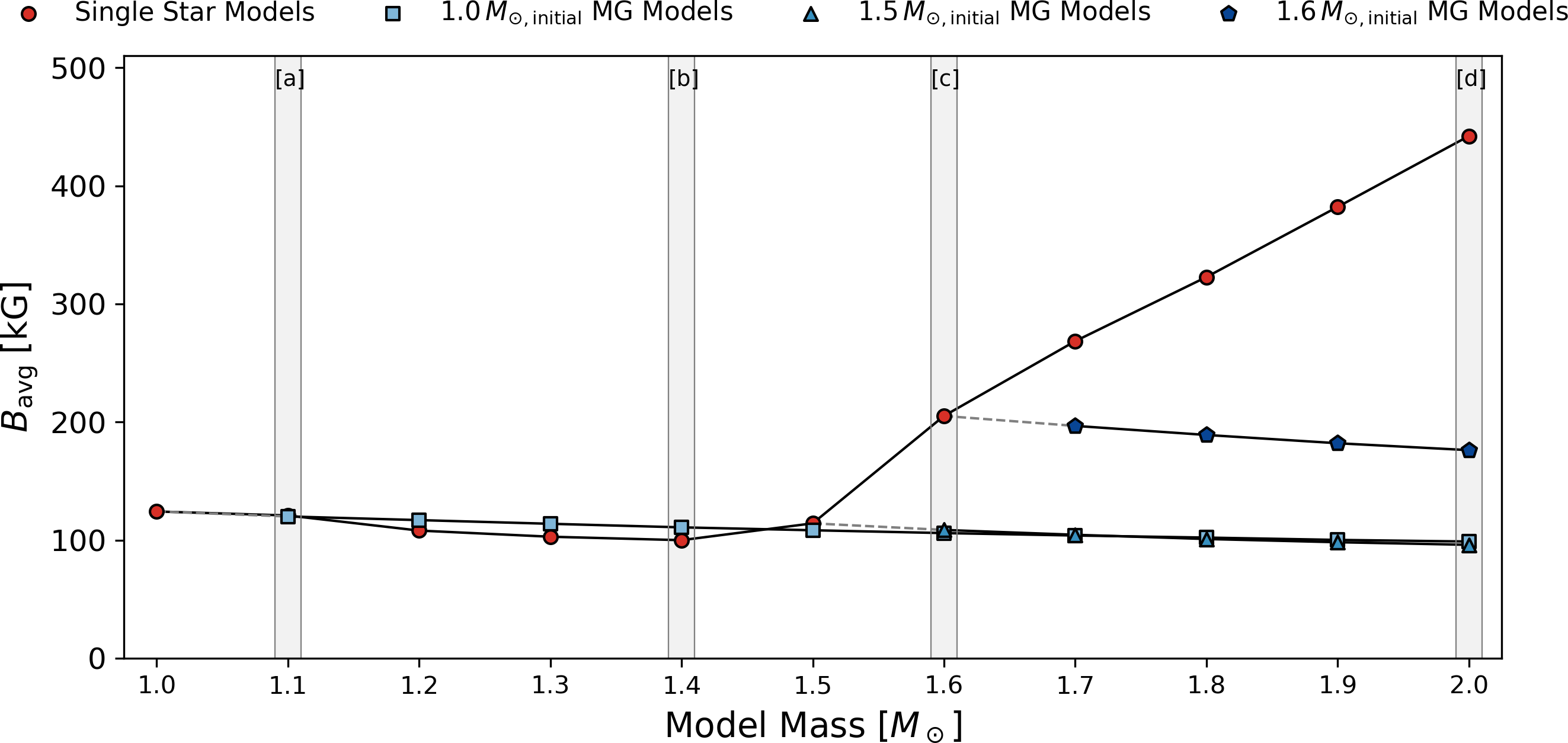}
    \caption{The weighted average of the radial magnetic field strength $B_\mathrm{avg}$ needed to produce a mean magnetic shift $\delta\nu_\mathrm{mag}$ of $0.2\,\mathrm{\mu Hz}$ at a frequency of $\nu_\mathrm{max}=180\,\mathrm{\mu Hz}$ for the single-star and mass-gain models. Here, the values for $B_\mathrm{avg}$ computed for the single-star models are signified by red dots and the values for the mass-gain models by light-blue squares ($M_\mathrm{init}=1.0\,M_\odot$), blue triangles ($M_\mathrm{init}=1.5\,M_\odot$), and dark-blue pentagons ($M_\mathrm{init}=1.6\,M_\odot$). The gray dashed lines connect the values of $B_\mathrm{avg}$ computed for the mass-gain models to the value of $B_\mathrm{avg}$ computed for the single-star model corresponding to their respective initial mass, while the gray areas mark the representative models (see Tab.~\ref{tab:modelproperties}).}
    \label{fig:avg_rad_B_field}
\end{figure*}

\begin{figure*}
    \sidecaption
    \includegraphics[width=12cm]{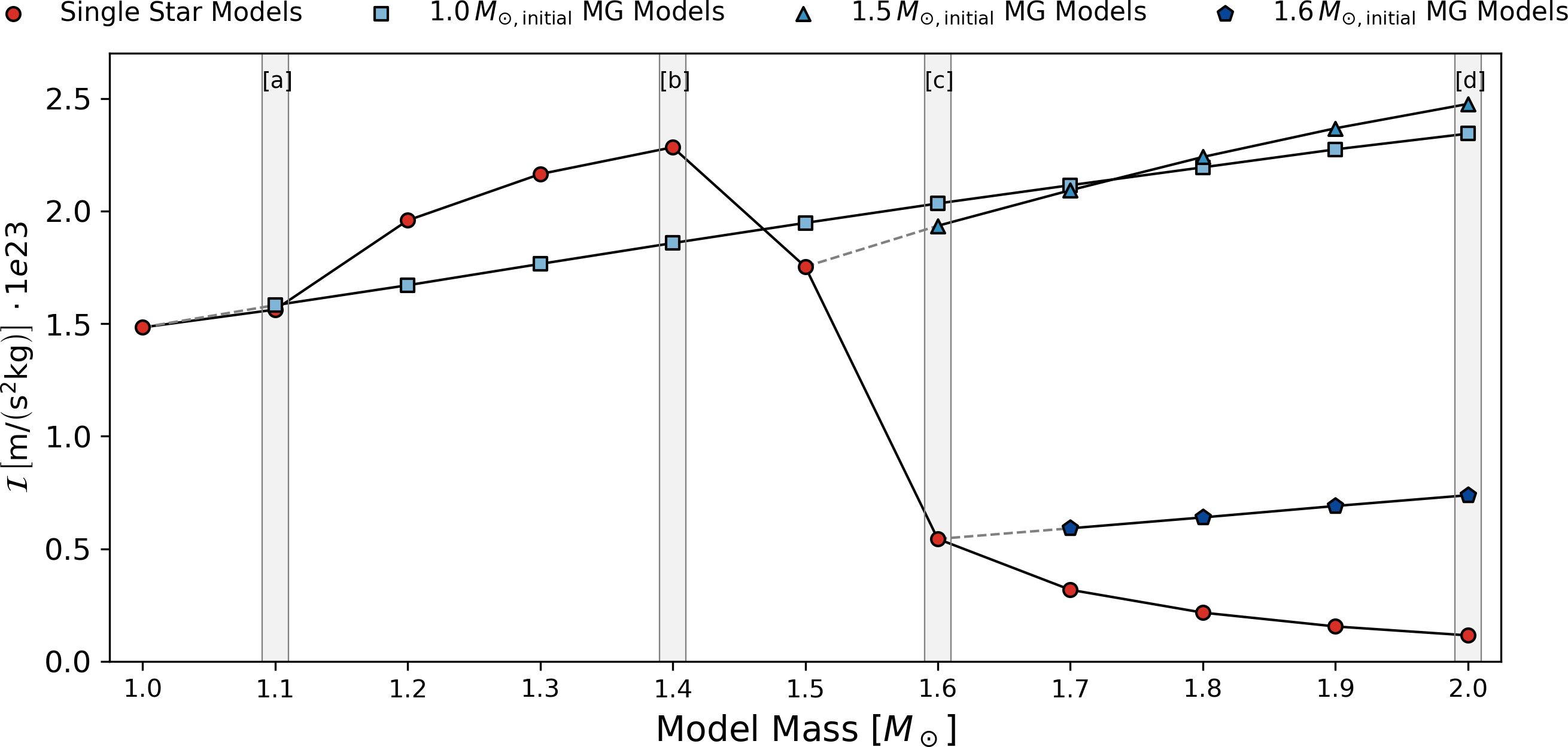}
    \caption{The global magnetic sensitivity $\mathcal{I}$ obtained for mass-gain and single-star models at a frequency value for the global power excess of $\nu_\mathrm{max}=180\,\mathrm{\mu Hz}$. The values for the global magnetic sensitivity $\mathcal{I}$ obtained for the single-star models are signified by red dots and the values for the mass-gain models by light-blue squares $\left(M_\mathrm{init}=1.0\,M_\odot\right)$, blue triangles $\left(M_\mathrm{init}=1.5\,M_\odot\right)$, and dark-blue pentagons $\left(M_\mathrm{init}=1.6\,M_\odot\right)$. The gray dashed lines connect the values of $\mathcal{I}$ obtained for the mass-gain models to the value of $\mathcal{I}$ of the single-star model corresponding to their respective initial mass, while the gray areas mark the values for $\mathcal{I}$ obtained for the representative models listed in Tab.~\ref{tab:modelproperties}.}
    \label{fig:structure_factor}
\end{figure*}

Internal magnetic fields perturb stellar oscillation frequencies, producing a measurable frequency shift $\delta\nu_\mathrm{mag}$ \citep{Das20,Bugnet21,Li22}. When only considering g-modes of high radial order $n$ in the asymptotic regime \citep[e.g.][]{Unno89}, it was shown \citep[see][]{Hasan05, Rashba07, Bugnet21} that the radial part of the magnetic field dominates the total influence on the frequency shift $\delta\nu_\mathrm{mag}$. The frequency shift can then be expressed as \citep{Li22},
\begin{equation}
    \label{eq:approx_mean_shift}
    \delta\nu_\mathrm{mag} = \frac{\mathcal{I}}{\left(2\pi\right)^4 \mu_0 \nu^3} \int_{r_{\rm{i}}}^{r_{\rm{o}}} K(r)\overline{B_r}(r)^2\,\mathrm{d}r,
\end{equation}
as long as the field has a significant radial component $B_r$ \citep{Mathis21, Bhattacharya24}. In Eq.~\ref{eq:approx_mean_shift}, $\mu_0$ is the vacuum permeability, and $r_{\rm{i}}$ and $r_{\rm{o}}$ are the inner and outer boundaries of the g-mode cavity respectively, the parameters $\mathcal{I},\,\nu$ and $\delta\nu_\mathrm{mag}$ are defined and discussed below. 

Equation~\ref{eq:approx_mean_shift} contains several quantities that are important to be defined properly in the context of this work. First and foremost, $\nu$ is the unperturbed eigenfrequency of the considered g-mode or g-dominated mixed mode. While $\nu$ is often directly associated with $\nu_\mathrm{max}$ in this specific context \citep[e.g. ][]{Li23, Villate26}, it originally refers to the eigenfrequency of a specific g-dominated mixed mode. The function $K$ is the asymptotic magnetic kernel sensitivity function \citep{Li22},
\begin{equation}
    K(r)=\frac{\displaystyle\frac{1}{\rho}\left(\frac{N}{r}\right)^3}{\displaystyle\int_{r_{\rm{i}}}^{r_{\rm{o}}}\left(\frac{N}{r}\right)^3\,\frac{\mathrm{d}r}{\rho}},
    \label{eq:magnetic_kernel}
\end{equation}
which quantifies the sensitivity of the oscillation mode to the magnetic field at the radius coordinate $r$, with $N$ and $\rho$ being the Brunt-Väisälä frequency and the density of the stellar medium, respectively. The quantity $\overline{B_r}$ in Eq.~\ref{eq:approx_mean_shift} represents the horizontally averaged radial magnetic field strength \citep{Li22}:
\begin{equation}
    \overline{B_r}(r)^2=\frac{1}{4\pi}\iint B_r(r,\theta,\varphi)^2\sin\theta\,\mathrm{d}\theta \, \mathrm{d}\varphi\, .
    \label{eq:mean_horizontal_field}
\end{equation}
Finally, the global magnetic sensitivity $\mathcal{I}$, referred to as the core structure factor in previous studies, is defined as \citep{Li22},
\begin{equation}
    \mathcal{I}=\frac{\displaystyle\int_{r_{\rm{i}}}^{r_{\rm{o}}}\left(\frac{N}{r}\right)^3\frac{\mathrm{d}r}{\rho}}{\displaystyle\int_{r_{\rm{i}}}^{r_{\rm{o}}}\left(\frac{N}{r}\right)\,\mathrm{d}r},
\label{eq:struc_fac}
\end{equation}
where $\rho$ is the density of the stellar medium at the radius coordinate $r$. The parameter $\mathcal{I}$ can be interpreted as the overall sensitivity of the oscillation modes to the magnetic field and can thus be seen as the global counterpart of $K$. Throughout this paper, we refer to $\mathcal{I}$ as the global magnetic sensitivity. 

The integral in Eq.~\ref{eq:approx_mean_shift} defines an average of the radial magnetic field strength weighted by the magnetic kernel sensitivity function \citep{Li22},
\begin{equation}
    \langle B_r^2\rangle = \int_{r_{\rm{i}}}^{r_{\rm{o}}} K(r)\overline{B_r}(r)^2\,\mathrm{d}r,
    \label{eq:Bavg_integral}
\end{equation}
which allows for the definition of the following expression,
\begin{equation}
    \label{eq:asymptotic_mag_field_strength}
    \langle B_r^2 \rangle=\frac{\left(2\pi\right)^4 \mu_0 \delta\nu_\mathrm{mag}\nu^3}{\mathcal{I}}.
\end{equation}
Equation~\ref{eq:asymptotic_mag_field_strength} provides an estimate of the weighted average of the radial magnetic field strength without requiring the assumption of a specific field geometry. In practice, the parameters $\delta\nu_\mathrm{mag}$ and $\nu$ are obtained from asteroseismic measurements, while the global magnetic sensitivity $\mathcal{I}$ must be inferred using stellar models. The construction of a representative model is thus crucial to properly constrain $\mathcal{I}$ and thus $\langle B_r^2\rangle$. Throughout the rest of this work, we will use the notation $B_\mathrm{avg}=\sqrt{\langle B_r^2\rangle}$ to refer to the weighted average of the radial magnetic field strength. 

As single-star and mass-gain stars of identical mass can possess different internal structures \citep{Rui21}, we expect to find different values for the global magnetic sensitivity $\mathcal{I}$. As a result, we further expect different values for $B_\mathrm{avg}$ for the two assumed evolutionary histories. In a similar manner, we expect that the asymptotic magnetic sensitivity kernel $K$ may differ between a mass-gain and single-star evolutionary history at identical total masses. 

\subsection{Impact of mass gain on magnetic frequency shifts}

\begin{figure*}[t]
    \centering
    \includegraphics[scale=0.63]{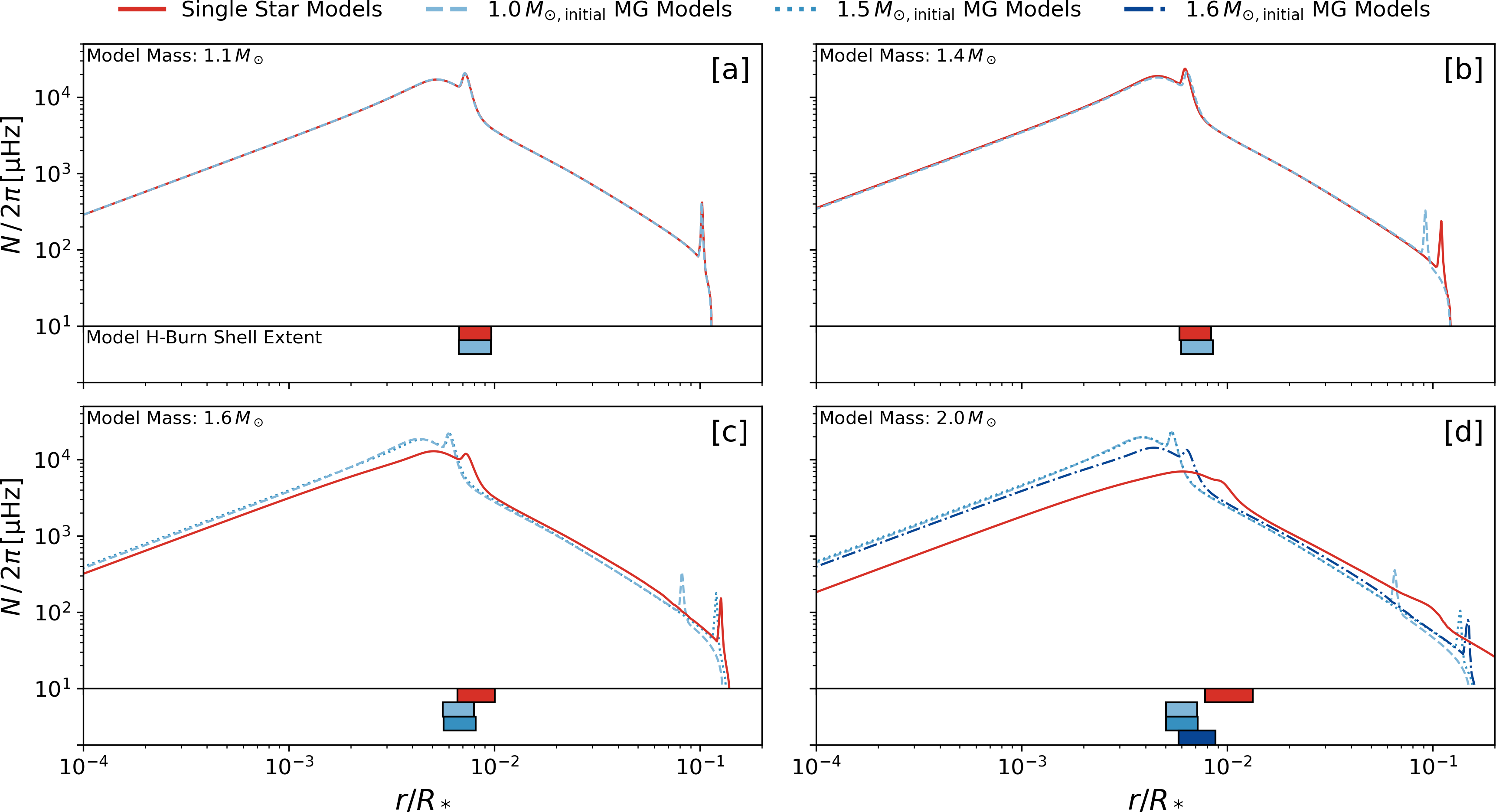}
    {\caption{The Brunt-Väisälä frequency as well as the model-specific extent of the hydrogen burning shell obtained for the representative single-star and mass-gain models at a value for the frequency of the global power excess of $\nu_\mathrm{max}=180\,\mathrm{\mu Hz}$. The colored boxes at the bottom of each panel signify the extent of the hydrogen-burning shell of the respective single-star or mass-gain model. The labels $\left[a\right]$, $\left[b\right]$, $\left[c\right]$, and $\left[d\right]$ correspond to the model labeling introduced in Tab.~\ref{tab:modelproperties}.}\label{BruntVas}}
\end{figure*}

We compared the values of $B_\mathrm{avg}$ obtained from the single-star and mass-gain models to study how a mass-gain event influences the sensitivity of mixed modes to the magnetic field. The following sections present the results of this investigation.

\subsubsection{Model and mass dependence of the inferred field strength}
\label{sec:magnetic_shifts}

For all of the models, we uniformly assume a mean magnetic frequency shift of $\delta\nu_\mathrm{mag}=0.2\,\mathrm{\mu Hz}$ for an assumed g-dominated dipole mixed mode. We take the frequency of this assumed g-dominated mixed mode as identical to the model frequency of $\nu_\mathrm{max}$. The value for $\delta\nu_\mathrm{mag}$ that we consider thus represents the frequency shift fitted at $\nu_\mathrm{max}$. Due to our extraction criterion in $\nu_\mathrm{max}$ stated in Sect.~\ref{sec:model_info}, we consider a mode frequency of $\nu=180\,\mu Hz$ for all our models. The assumed frequency value for the magnetic shift falls well into the measured frequency range for $\delta\nu_\mathrm{mag}$ \citep[e.g., the value reported for KIC 8684542 in][]{Li22}. By calculating $\mathcal{I}$ from each MESA profile, we then inferred the value of $B_\mathrm{avg}$ via Eq.~\ref{eq:asymptotic_mag_field_strength}. This approach can also be interpreted as finding the magnetic field strength necessary to cause the magnetic shift. In this sense, it is possible to establish a clear connection between the value of $B_\mathrm{avg}$ and the mode sensitivity to the magnetic field, with lower values of $B_\mathrm{avg}$ corresponding to higher sensitivities and vice versa. We show the resulting values of $B_\mathrm{avg}$ in Fig.~\ref{fig:avg_rad_B_field}.

Since the magnetic frequency shift is fixed at the same value for all models, any difference in $B_\mathrm{avg}$ is purely due to a model-dependent difference in the global sensitivity parameter $\mathcal{I}$, which is closely related to the properties of the model radiative core (see Eq.~\ref{eq:struc_fac}).

For the single-star models (red points), Fig.~\ref{fig:avg_rad_B_field} shows a clear two-regime behavior in mass, with the transition occurring between $1.5$ and $1.6\,M_\odot$. We found that the transition mass is sensitive to $\nu_\mathrm{max}$ and shifts towards lower masses for larger $\nu_\mathrm{max}$ values and to larger masses for lower $\nu_\mathrm{max}$ values. Single-star models with masses below the transition mass produce the same frequency shift at lower $B_\mathrm{avg}$, consistent with the findings of \cite{Buchele26}, indicating a higher sensitivity to the magnetic field compared to their counterparts with masses above the transition mass.

For our mass-gain models (blue symbols in Fig.~\ref{fig:avg_rad_B_field}), no such transition is present. Instead, $B_\mathrm{avg}$ shows a slight decrease with increasing total model mass for models with $M_\mathrm{init}=1.0\,M_\odot$ and $M=1.5\,M_\odot$. Models with $M_\mathrm{init}=1.6\,M_\odot$ are an exception, showing a higher $B_\mathrm{avg}$ and thus lower sensitivity to the magnetic field than mass-gain models with initial masses of $M_\mathrm{init}=1.0\,M_\odot$ and $1.5\,M_\odot$. Above a total mass of $\approx 1.5\,M_\odot$, all mass-gain models show a lower $B_\mathrm{avg}$ and thus higher sensitivity to the field than the single-star models of corresponding mass.

\subsubsection{Dependency on stellar core structure and the global magnetic sensitivity}
\label{sec:stellar_structure}

The behavior in Fig.~\ref{fig:avg_rad_B_field} can be explained by the differences in the global magnetic sensitivity of mass-gain and single-star models depicted in Fig.~\ref{fig:structure_factor}, which itself arises due to differences in the structure of the model cores. As the effect exerted by a mass-gain event on the structure of the radiative model core is negligible \citep{Rui21}, mass-gain models possess a core that is very similar to that of an RGB single-star model whose mass is identical to their initial mass. The global magnetic sensitivity is mainly determined by the Brunt-Väisälä profile model (see Eq.~\ref{eq:struc_fac}), which is conserved during the mass-gain event, as visible in Fig.~\ref{BruntVas}. Another parameter capable of influencing the value of the global magnetic sensitivity is the model density $\rho$, as is evident from Eq.~\ref{eq:struc_fac}. We find, that the model density $\rho$ is roughly conserved during the mass-gain event. This causes a significant difference in the $\rho$ profile of mass-gain and single-stars at identical total masses $M_\mathrm{total}>1.5\,M_\odot$, as visible in Fig.~\ref{fig:internal_density}. As a result, mass-gain models will show a similar value for $\mathcal{I}$, and thus $B_\mathrm{avg}$, before and after the mass-gain event, as their $N$ and $\rho$ profiles barely change. 

\begin{figure}[t]
    \centering
    \includegraphics[scale=0.61]{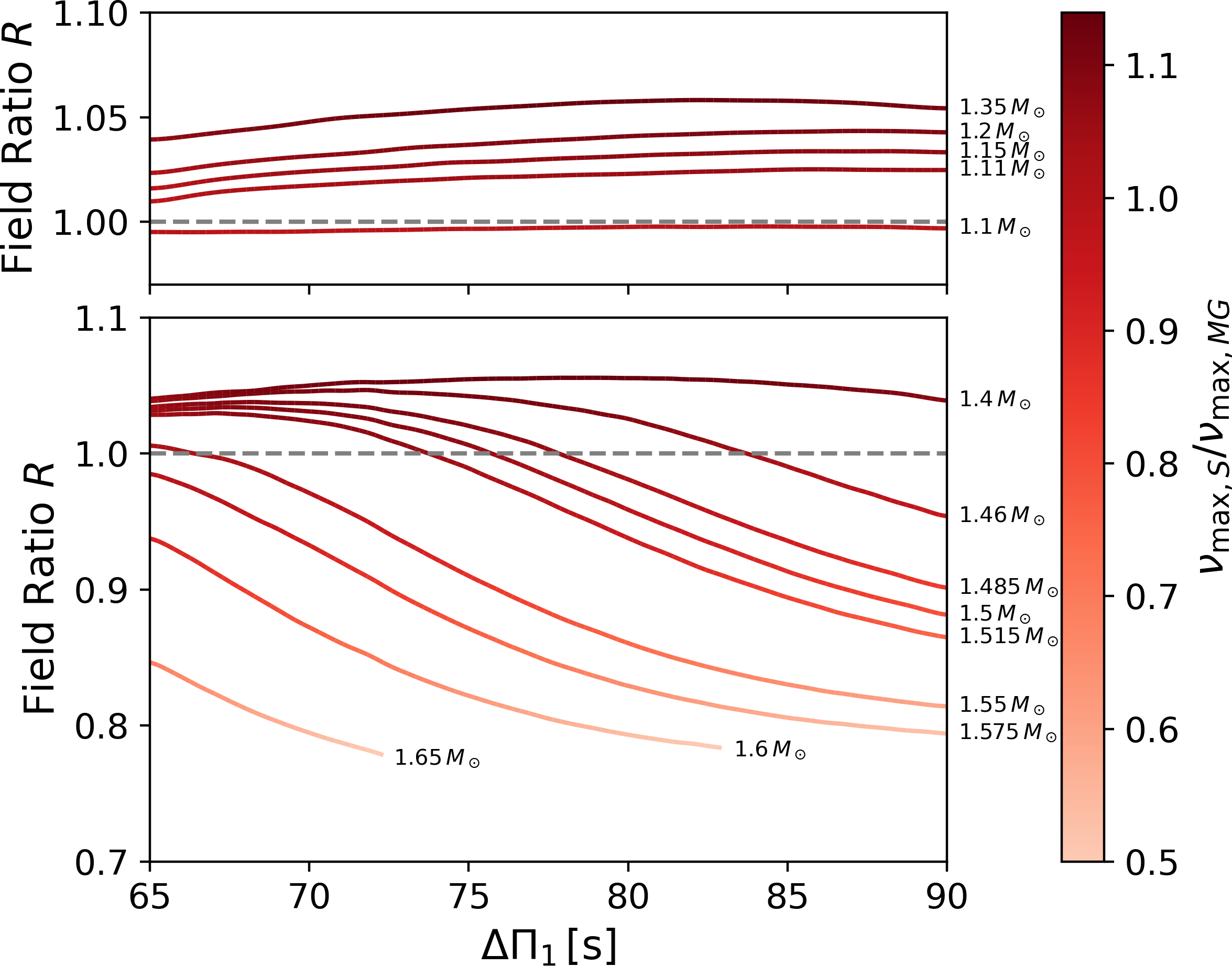}
    {\caption{The ratio $R_\mathrm{B}$ of the weighted average of the radial magnetic field strength $B_\mathrm{avg}$ of the mass-gain and single-star models needed to produce a magnetic shift of $0.2\,\mathrm{\mu Hz}$ for a parameter space spanned by the asymptotic period spacing of dipole g-modes $\Delta\Pi_1\in\left[65\,\mathrm{s},\,90\,\mathrm{s}\right]$ and total model mass $M\in\left[1.1\,M_\odot,\,2.0\,M_\odot\right]$. The different tracks for $R_\mathrm{B}$ are color-coded according to the $\nu_\mathrm{max}$ ratio of the compared models. Dashed lines indicate $\Delta\Pi_1-R_\mathrm{B}$ for which the model difference in $\nu_\mathrm{max}$ is greater than $50\%$. The top panel shows the $R_\mathrm{B}$ tracks for a mass regime of $M\in \left[1.1\,M_\odot ,\, 1.35\,M_\odot \right]$ and the bottom panel the $R_\mathrm{B}$ tracks for a mass regime of $M\in \left[1.4\,M_\odot , \, 2.0\, M_\odot \right].$}\label{fig:field_ratio_evo}}
\end{figure}

Looking at Fig. ~\ref{fig:structure_factor}, the increase of $\mathcal{I}$ with total mass $M_\mathrm{total}$ for mass-gain models of identical initial mass $M_\mathrm{init}$ is a result of our profile extraction criterion in $\nu_\mathrm{max}$. During the assumed mass-gain event, the effective temperature, radius, and mass of the model are altered, which leads to an overall increase of  $\nu_\mathrm{max}$. The mass-gain event thus has a significant impact on the conditions in the model envelope. Mass-gain models that gain a lot of mass thus take longer to reach the extraction threshold of $\nu_\mathrm{max}=180\;\mathrm{\mu Hz}$ compared to models that gain less mass, allowing the cores of the latter to evolve for a longer time. This results in models that gain more mass, possessing more evolved cores, causing the constant increase of $\mathcal{I}$ with mass as visible in Fig.~\ref{fig:structure_factor}. This behavior is illustrated in Fig.~\ref{fig:numax_vs_sf}, in which we show the relationship between $\nu_\mathrm{max}$ and $\mathcal{I}$ under the consideration of a mass-gain event. For clarity, we show only mass-gain models with $M_\mathrm{init}=1.0\,M_\odot$ in Fig.~\ref{fig:numax_vs_sf}. However, the behavior is consistent for all mass-gain models. As all mass-gain models of identical $M_\mathrm{init}$ possess cores that mainly differ in their degree of evolution, it follows that the trend of $\mathcal{I}$ and $B_\mathrm{avg}$ with mass can be extended to also incorporate the single-star model of mass $M_\mathrm{init}$, as is indicated by the dashed lines in Fig.~\ref{fig:avg_rad_B_field} and Fig.~\ref{fig:structure_factor}.

In addition to $\mathcal{I}$, the behavior of the asymptotic sensitivity kernel function $K$ is of interest, as it represents the local sensitivity of oscillation modes to the magnetic field. In Fig.~\ref{fig:mag_kern}, we present the $K$ profiles as well as their cumulative profiles for the representative single-star and mass-gain models. The correlation of the peak of $K$ and the Brunt-Väisälä frequency \citep[e.g.][]{Li22} and emerges directly from the $K\propto N^3$ dependence visible in Eq.~\ref{eq:magnetic_kernel}. In the literature, it is common to treat $B_\mathrm{avg}$ as an estimate of the magnetic field strength near the hydrogen burning shell, owing to the sensitivity of $K$ to the Brunt-Väisälä frequency \citep[e.g.][]{Li22}. From Fig.~\ref{fig:mag_kern}, we find that this assumption is even more accurate for mass-gain models, as the cumulative profiles for $K$ show the largest increase in a very narrow region in terms of radius around the hydrogen burning shell at all total masses considered in the figure.

In summary, the model- and mass-dependent differences of $B_\mathrm{avg}$, $\mathcal{I}$, and $K$ that emerge for constant $\delta\nu_\mathrm{mag}$ and $\nu$ can be directly traced back to differences in the $N$ and $\rho$ profiles of the two model types at identical total masses that arise due to the mass-gain event. In particular, the mass-gain event changes the relationship between the properties of the model core and model envelope. Thus, any fitting method that assumes standard single-star evolution risks yielding an incorrect value for the global magnetic sensitivity and, consequently, a biased estimate of $B_\mathrm{avg}$ when performed for a mass-gain remnant.

\subsubsection{Evolutionary change of the magnetic field strength ratio $R$}
\label{sec:field_ratio_evo}

In order to compare mass-gain and single-star models throughout different evolutionary stages, we define the magnetic field strength ratio,
\begin{equation}
    R_\mathrm{B}=\frac{B_\mathrm{MG}}{B_\mathrm{S}} = \sqrt{\frac{\mathcal{I}_\mathrm{S}}{\mathcal{I}_\mathrm{MG}}},
    \label{eq:R_evo}
\end{equation}
where $B_S$ and $B_{MG}$ are the field strengths calculated for single-star and mass-gain models of identical total model mass. The definition of $R_\mathrm{B}$ allows us to visualize the relative evolution of $B_\mathrm{avg}$ for single-star and mass-gain models along a chosen evolutionary parameter.

Up to this point, we have used $\nu_\mathrm{max}$ to determine the point in stellar evolution at which we performed our seismic analysis on the models. Our choice was motivated by the requirement of having the frequency $\nu$ of the g-dominated mixed mode as close to $\nu_\mathrm{max}$ as possible. This ensures that the mixed mode and thus the magnetic frequency shift is indeed measurable at the considered frequency. However, while this partially constrains the stellar envelope conditions across single-star and mass-gain models, the core parameters are left virtually unconstrained. We now add an extraction parameter that is better suited to constrain the stellar core \citep{Deheuvels_mg22,Villate26}, the asymptotic period spacing of dipole g-modes \citep[e.g.][]{Aerts10},
\begin{equation}
    \label{eq:dpi}
    \Delta\Pi_1=\frac{2 \pi^2}{\sqrt2} \left(\int_{r_{\rm{i}}}^{r_{\rm{o}}}\frac{N}{r}\,\mathrm{d}r\right)^{-1}.
\end{equation}

We now compare models at multiple values of $\Delta\Pi_1$, in order to understand the evolutionary behavior of $R_\mathrm{B}$ at different model masses. We choose to study $R_\mathrm{B}$ for the mass-gain models with initial mass $M_\mathrm{init}=1.0\,M_\odot$ in a parameter space spanned by $\Delta\Pi_1\in \left[65\,\mathrm{s},\,90\,\mathrm{s}\right]$ and $M\in\left[1.1\,M_\odot,\,2.0\,M_\odot\right]$. The resulting evolution of $R_\mathrm{B}$ along $\Delta\Pi_1$ is shown in Fig.~\ref{fig:field_ratio_evo}.

Due to our change of the extraction parameter to $\Delta\Pi_1$, the values for $\nu_\mathrm{max}$ differ significantly between the compared models. This mismatch is problematic, as we still impose that $\delta\nu_\mathrm{mag}$ and $\nu$ are identical when calculating $R_\mathrm{B}$. If the values for $\nu_\mathrm{max}$ of the compared single-star and mass-gain models are too far apart, the observation of a mixed mode and, as a result, the measurement of $\delta\nu_\mathrm{mag}$, would not be feasible at identical $\nu$ in both models. To address this, we impose an observability condition on $\nu_\mathrm{max}$ by defining the ratio $R_\mathrm{\nu_{max}}=\frac{\nu_\mathrm{max,S}}{\nu_\mathrm{max,MG}}$. When $R_\mathrm{\nu_\mathrm{max}}\in\left[0.5,\;1.5\right]$, we consider mixed modes of identical frequency to be observable, indicated by solid lines in Fig.~\ref{fig:field_ratio_evo}. Outside this range, we treat them as unobservable simultaneously, and they are, as a result, not included in Fig.~\ref{fig:field_ratio_evo}. 

From Fig.~\ref{fig:field_ratio_evo}, we find that different total model masses result in distinct trends of the $\Delta\Pi_1-R_\mathrm{B}$ curve. We find that $R_\mathrm{B}$ ranges from  1 to 1.05 when the total model mass is below $1.4\,M_\odot$ with little dependence on $\Delta\Pi_1$ (see upper panel of Fig.~\ref{fig:field_ratio_evo}). The mass range $1.4\,M_\odot< M < 1.6\,M_\odot$ marks a transition region in which $R_\mathrm{B}$ changes with $\Delta\Pi_1$, showing an increase with decreasing $\Delta\Pi_1$. The difference in the $R_\mathrm{B}$ tracks is a direct result of the structural differences of the model core owing to the assumed mass-gain event (see Sect.~\ref{sec:stellar_structure}).

In summary, we find that the difference in $B_\mathrm{avg}$ inferred for mass-gain and single-star models at identical values for $\Delta\Pi_1$ is up to $25\%$ in the considered $\Delta\Pi_1-M$ parameter space when considering modes that are observable in both single-star and mass-gain models. We furthermore find that in the considered $\Delta\Pi_1$ range, all models with total masses $>1.65\,M_\odot$ cannot fulfill our observability criterion.

\subsection{$B_\mathrm{avg}$ estimates for possible magnetic merger remnants}

Having established the impact of a mass-gain event on the sensitivity of stellar oscillations to the magnetic field, we now apply these findings to two RGB stars that have been identified as possible merger candidates in the literature. The recent work of \cite{Villate26} reports the magnetic field strength for several seismically active RG stars, including KIC 4350501, which was already identified as a merger candidate by \cite{Deheuvels_mg22} because its $\Delta\nu$ and $\Delta\Pi_1$ values place it well below the degeneracy sequence \citep[see ][]{Deheuvels_mg22} in the $\Delta\nu$-$\Delta\Pi_1$ diagram. From the magnetic RG sample of \cite{Hatt24}, we identified KIC 9508757, which also exhibits a $\Delta\nu/\Delta\Pi_1$ combination inconsistent with standard single-star evolution. In this section, we demonstrate that it is possible to construct acceptable models for both KIC 4350501 and KIC 9508757 and to constrain measures for the magnetic field strength under the assumption of a mass-gain history.

\subsubsection{Definition of the stellar model grid}

As our original model grid assumes a constant metallicity for all models and the resolution of the initial- and post-gain mass parameters is insufficient, we computed a new mass-gain model grid consisting of 1320 model tracks. The mass and metallicity ranges were chosen to encompass the literature values for mass \citep{Yu18} and metallicity \citep{Mathur17} for KIC 4350501 and KIC 9508757. We varied the models in initial mass ($M_\mathrm{init}\in\left[1.2\,M_\odot,\,1.45\,M_\odot\right],\,\Delta M_\mathrm{init}=0.05\,M_\odot$), post gain mass ($M\in\left[1.5\,M_\odot,\,2.0\,M_\odot\right],\,\Delta M=0.05\,M_\odot$), initial metallicity ($\left[\mathrm{Fe/H}\right]\in\left[-0.36\,\mathrm{dex},\,-0.08\,\mathrm{dex}\right],\,\Delta\left[\mathrm{Fe/H}\right]=0.07\,\mathrm{dex}$), and the mixing length parameter ($\alpha_\mathrm{MLT}\in\left[2.0,\,2.6\,\right],\,\Delta \alpha_\mathrm{MLT}=0.2$). To test whether we could find a fitting model assuming a conventional single-star evolution, we computed an additional grid consisting of 300 single-star models. The models vary in initial mass ($M_\mathrm{init}\in\left[1.3\,M_\odot,\,2.0\,M_\odot\right],\,\Delta M_\mathrm{init}=0.05\,M_\odot$), metallicity and the mixing length parameter, with grid points identical to those of the mass-gain model grid. %In order to not bias the grids, we also varied $\alpha_\mathrm{MLT}$ for the single-star model grid.  

\subsubsection{Fitting of the model grids}

We identified the best fitting model by calculating $\chi^2_\mathrm{fit}$ following the same approach as \cite{Villate26},
\begin{equation}
\chi^2_\mathrm{fit}=\frac{\left(\Delta\nu_\mathrm{o}-\Delta\nu_\mathrm{m}\right)^2}{\sigma_\mathrm{\Delta\nu}^2} + \frac{\left(\Delta\Pi_\mathrm{1,o}-\Delta\Pi_{\mathrm{1,m}}\right)^2}{\sigma_{\Delta\Pi_1}^2} +\frac{\left(\nu_\mathrm{max,o}-\nu_\mathrm{max,m}\right)^2}{\sigma_\mathrm{\nu_{max}}^2},
\end{equation}
where the subscript $m$ refers to model values, and the subscript $o$ to observational values taken from \cite{Hatt24}, and \cite{Villate26} ($\Delta\Pi_1$ and $\Delta\nu$), and \cite{Yu18} ($\nu_\mathrm{max}$), with $\sigma_i$ being the corresponding stated uncertainties. We adopt this definition as the parameters $\Delta\nu, \nu_\mathrm{max}$ and $\Delta\Pi_1$ jointly constrain both the core and envelope parameters of the model. Constraining only the envelope or the core leads to large uncertainties in $\mathcal{I}$, as shown in Sect.~\ref{sec:stellar_structure} and Sect.~\ref{sec:field_ratio_evo}. We expect that constraining both regions simultaneously reduces these uncertainties.

\begin{figure}[t]
    \centering
    \includegraphics[scale=0.53]{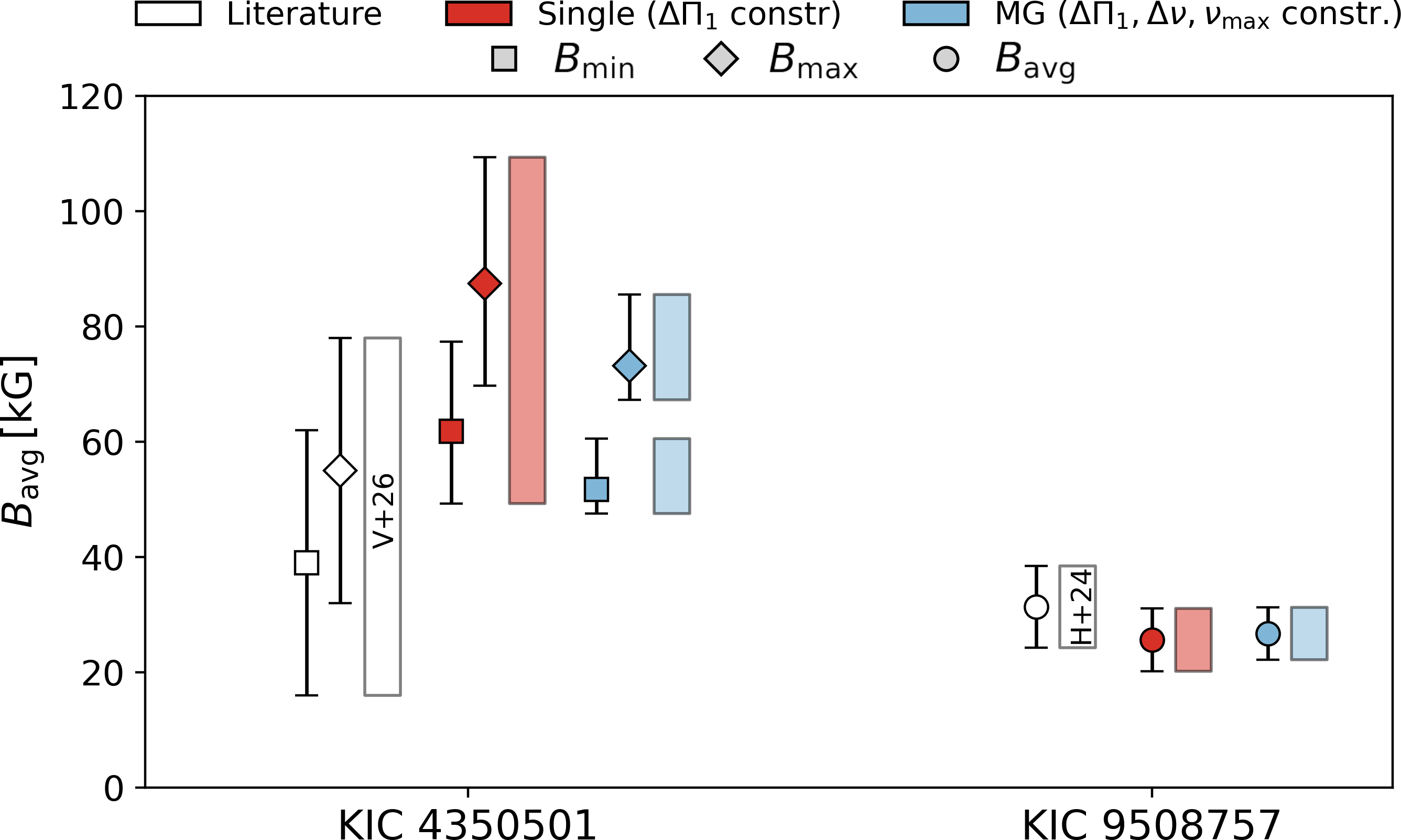}
    {\caption{Comparison of the values for $B_\mathrm{avg}$ (circles) for KIC 9508757 and $B_\mathrm{min}$ (squares) and $B_\mathrm{max}$ (diamonds) for KIC 4350501 yielded from our mass-gain and single-star model grid and the corresponding literature values reported in \cite{Hatt24} (Indicated by the H+24 label) and \cite{Villate26} (Indicated by the V+26 label) respectively. The values obtained for the best-fitting models are indicated by markers, while the total uncertainty range is indicated by colored rectangles.}
    \label{fig:merger_targets_b}}
\end{figure}

We calculate $\Delta\nu$ using the radial modes found by the GYRE stellar oscillation code \citep{Townsend13} in the five radial orders closest to $\nu_\mathrm{max}$. We define $\Delta\nu$ to be the slope of a linear fit of the corresponding $\ell=0$ mode frequencies \citep[for the methodology e.g.,][]{Grossmann25}. To mitigate the effects of coarse evolutionary resolution, we linearly interpolated $\Delta\Pi_1,\nu_\mathrm{max}$ and $\Delta\nu$ onto a grid a thousand times finer in model age. 

\begin{figure*}[t]
\centering
\begin{minipage}[c]{0.68\textwidth}
    \centering
    \includegraphics[width=\linewidth]{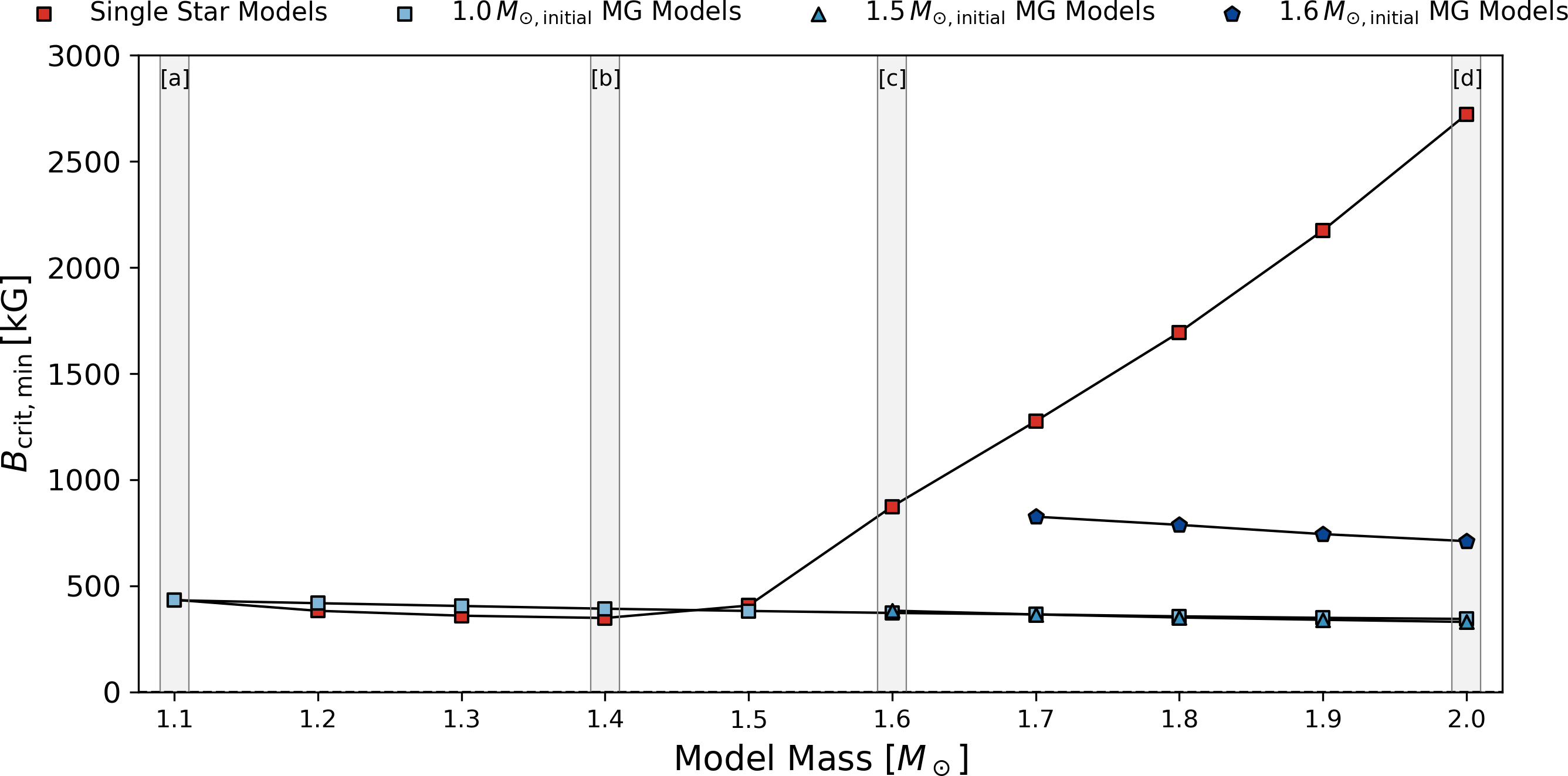}
\end{minipage}\hfill
\begin{minipage}[c]{0.32\textwidth}
    \captionsetup{type=figure, labelfont=bf, font=small,
                  justification=raggedright, singlelinecheck=false}
    \captionof{figure}{The minimum critical magnetic field strength $B_\mathrm{crit,min}$ required to cause suppression of the mixed dipole modes in the mass-gain and single-star models at a frequency of $\nu=\nu_\mathrm{max}=180\,\mathrm{\mu Hz}$. The values for $B_\mathrm{crit,min}$ yielded for the single-star models are signified by red dots while the values for $B_\mathrm{crit,min}$ for the mass-gain models are signified by light-blue squares $\left(M_\mathrm{init}=1.0\,M_\odot\right)$, blue triangles $\left(M_\mathrm{init}=1.5\,M_\odot\right)$, and dark-blue pentagons $\left(M_\mathrm{init}=1.6\,M_\odot\right)$. The gray areas mark the yielded values for the representative models (see Tab.~\ref{tab:modelproperties}).}
    \label{fig:B_crit}
\end{minipage}
\end{figure*}

In the case of KIC 4350501, we find three mass-gain models for which $\chi^2_\mathrm{fit}<2$. In order to find a fitting model for KIC 9508757, we loosened the acceptance threshold to $\chi^2_\mathrm{fit}<6$, which is fulfilled by four mass-gain models. Although only a small number of mass-gain models satisfy our fitting criteria, we consider it sufficient for our proof-of-concept approach, as a detailed fitting of these targets is beyond the intended scope of this work.

For the single-star model grid, we were, as in the literature, not able to find an acceptable model. The best-fitting single-star models show $\chi^2_\mathrm{fit}\approx10^3-10^4$ in the case of both KIC 4350501 and KIC 9508757. For comparison purposes, we employ the loosened fitting prescription used by \cite{Villate26}, which solely constrains $\Delta\Pi_1$,
\begin{equation}
    \label{eq:chi_fit_single}
    \chi^2_\mathrm{fit,s}= \frac{\left(\Delta\Pi_\mathrm{1,o}-\Delta\Pi_{\mathrm{1,m}}\right)^2}{\sigma_{\Delta\Pi_1}^2} ,
\end{equation}
for our single-star model grid. We slightly tighten the new fitting criterion by additionally demanding that fitting models must fulfill the observability condition in $\nu_\mathrm{max}$ defined in Sect.~\ref{sec:field_ratio_evo}, rejecting all models that show a $\nu_\mathrm{max,m}$ deviating by more than $50\%$ from $\nu_\mathrm{max,o}$. This constrains the core properties but only very loosely constrains the envelope properties for the single-star models. We thus expect the model uncertainties discussed in Sect.~\ref{sec:field_ratio_evo} to emerge. Applying the fitting criterion defined in Eq.~\ref{eq:chi_fit_single}, we were able to find an abundance of single-star models fulfilling $\chi^2_\mathrm{fit}<2$. 

\subsubsection{Magnetic field estimates}

By calculating $\mathcal{I}$ from the fitting single-star and mass-gain models and assuming the values for $\delta\nu_\mathrm{mag}$ reported in \cite{Hatt24} and \cite{Villate26}, we estimated $B_\mathrm{avg}$ following Eq.~\ref{eq:asymptotic_mag_field_strength} for KIC 9508757. However, due to the doublet nature of the $\ell=1$ mixed modes of KIC 4350501, \cite{Villate26} cannot constrain the asymmetry parameter $a$. As a result, they obtain an interval of possible field strengths bounded by $B_\mathrm{min}$ and $B_\mathrm{max}$, which are defined as \citep{Villate26},
\begin{equation}
    B_\mathrm{min}=\left(\frac{2}{3}\frac{\left(2\pi\right)^4\mu_0 \delta\nu_\mathrm{mag}\nu^3}{\mathcal{I}}\right)^{1/2} \;,\; B_\mathrm{max}=\left(\frac{4}{3}\frac{\left(2\pi\right)^4\mu_0 \delta\nu_\mathrm{mag}\nu^3}{\mathcal{I}}\right)^{1/2},
    \label{eq:bmin_bmax}
\end{equation}
which we also adopt in the case of KIC 4350501. 

We estimate the uncertainty in $B_\mathrm{avg}$, $B_\mathrm{min}$, and $B_\mathrm{max}$ from the reported uncertainties in $\delta\nu_\mathrm{mag}$ and $\nu_\mathrm{max}$. For $\mathcal{I}$, we adopt a model uncertainty equal to half the range spanned by the values obtained from the fitting models. For comparison, the uncertainty implied by our single-star model grid corresponds to $27\%$ of the mean value of $\mathcal{I}$ across all fitting single-star models and is in good agreement with the value uncertainty reported by \cite{Hatt24}. We compare our values to the previously reported measurements.

\subsubsection{Discussion and comparison with literature}

The inferred values for $B_\mathrm{min},\,B_\mathrm{max}$, and $B_\mathrm{avg}$ calculated for the best-fitting single-star and mass-gain models compared to the values found by \cite{Hatt24} and \cite{Villate26} are shown in Fig.~\ref{fig:merger_targets_b}. While the $B_\mathrm{avg}$ estimate for KIC 9508757 is broadly consistent with that of \cite{Hatt24}, we find a significant offset relative to the value reported by \cite{Villate26} for KIC 4350501. As this offset in $B_\mathrm{min}$ and $B_\mathrm{max}$ appears for both the single-star and mass-gain results, we attribute it to differences in the stellar model prescriptions used by us and those used by \cite{Villate26}, an effect first discussed in \citep{Buchele26}. 

The ranges of $B_\mathrm{min}$, and $B_\mathrm{max}$ spanned by the fitting mass-gain models in the case of KIC 4350501 are significantly smaller than those obtained from the fitting single star models. While the difference in $B_\mathrm{avg}$ is less pronounced in the case of KIC 9508757, the range spanned by the fitting mass-gain models is still smaller compared to the range spanned by the fitting single-star models. As the envelope is nearly unconstrained for the fitting single-star models, the effects discussed in Sect.~\ref{sec:field_ratio_evo} cause a spread in $B_\mathrm{min},\,B_\mathrm{max}$, and $B_\mathrm{avg}$. In contrast, fitting both $\Delta\nu$ and $\nu_\mathrm{max}$ additionally to $\Delta\Pi_1$ in the mass-gain models also tightly constrains the model envelope and therefore reduces model ambiguity. 

The model parameters we obtain for $\Delta\Pi_1,\Delta\nu$, mass, radius, and metallicity in the case of the fitting mass-gain models agree well with the values reported in literature \citep{Mathur17, Yu18, Hatt24,Villate26}, however, we find that the effective temperature $T_\mathrm{eff}$ of the model stars is significantly higher than the observational values reported in \cite{Yu18}, especially in the case of KIC 4350501 ($5\sigma$ difference between $T_\mathrm{eff,o}$ and $T_\mathrm{eff,m}$ for KIC 4350501 and $2\sigma$ difference for KIC 9508757). We also find that $\nu_\mathrm{max}$ lies within $3\sigma$ of the observational values reported by \cite{Yu18} in the case of KIC 9508757, which explains the necessity of relaxing the $\chi^2$ fitting threshold to a value of $6$. We attribute this offset to the toy-model mass-gain approach we are using, and recommend that mass accretion be carefully simulated when measuring magnetic field strength on potential merger products. Due to the discrepancy in $T_\mathrm{eff}$ and $\nu_\mathrm{max}$, we emphasize that the parameter intervals for $B_\mathrm{min},\,B_\mathrm{max}$, and $B_\mathrm{avg}$ calculated by us are not necessarily more accurate than the values reported in \cite{Villate26} and \cite{Hatt24}. We do, however, argue that, for a self-consistent model prescription, the assumption of a mass-gain scenario allows $B_\mathrm{min},\,B_\mathrm{max}$, and $B_\mathrm{avg}$ to be more tightly constrained as it allows for the simultaneous constraining of the model envelope and model core.

\begin{figure*}[t]
    \centering
    \includegraphics[scale=0.6]{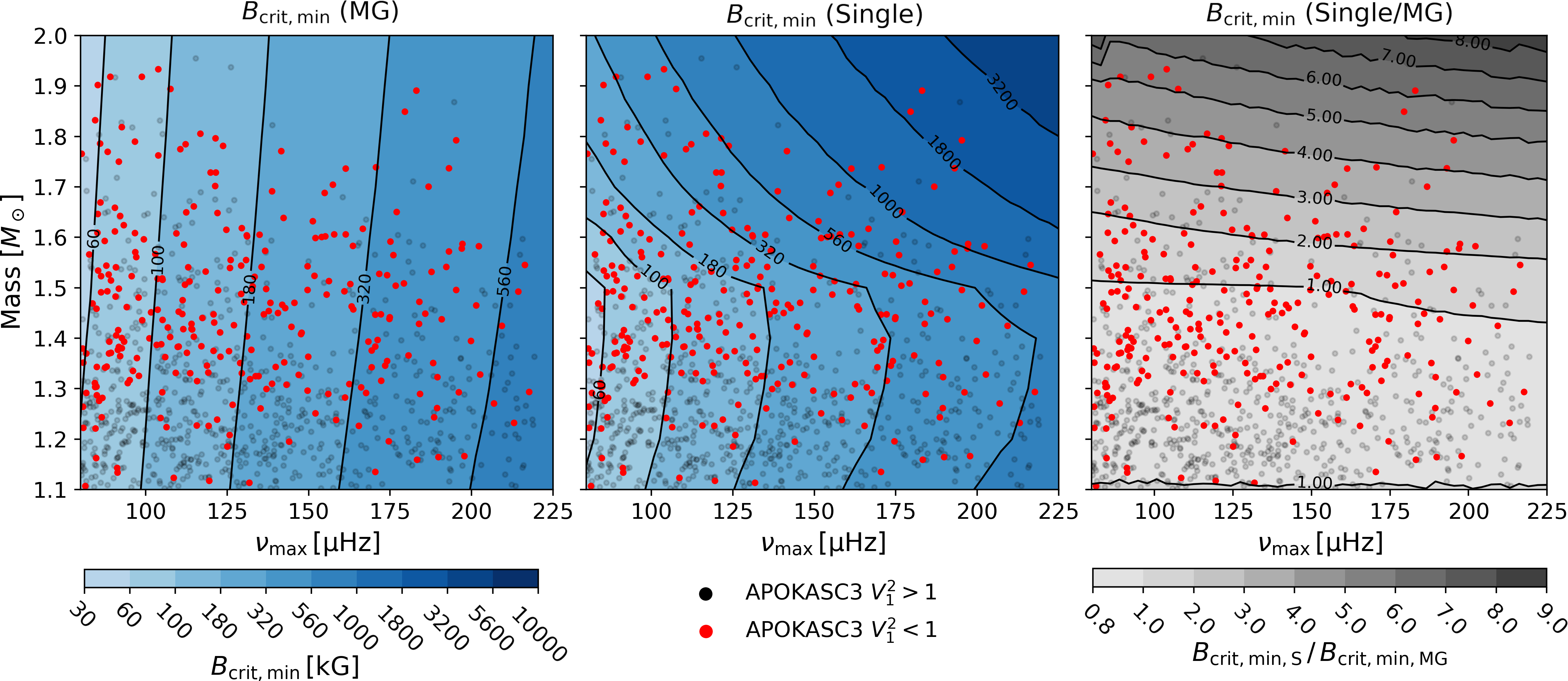}
    {\caption{The parameter space taken on by $B_\mathrm{crit,min}$ under the assumption of a $M_\mathrm{init}=1.0\,M_\odot$ mass-gain evolutionary history (left) and a single-star evolutionary history (middle), as well as the ratio of $B_\mathrm{crit,min}$ corresponding to the $M_\mathrm{init}=1.0\,M_\odot$ mass-gain or single-star assumption (right). The labeled curves in the background of the plots correspond to the contour lines of $B_\mathrm{crit,min}$ (left and middle panel) and the $B_\mathrm{crit,min}$ ratio between mass gain and single stat models (right panel) at different mass and $\nu_\mathrm{max}$ combinations. Targets that are listed in the APOKASC3 catalog as possessing low dipole mode visibilities are indicated by red dots, while targets that show normal dipole mode visibilities are indicated by transparent black dots.}\label{fig:depressed_mg_scenario}}
\end{figure*}

In summary, for a weak magnetic field the oscillations of mass-gain models with initial masses $\leq1.6\,M_\odot$ show a greater sensitivity to the magnetic field than those of single-star models above a total mass threshold of $\gtrsim1.5 M_\odot$ when the mass gain event is simulated during the early RGB evolutionary phase and the models are evaluated at $\nu_\mathrm{max}=180\,\mathrm{\mu Hz}$ (see Fig.~\ref{fig:structure_factor}). This exact value of the threshold depends on $\nu_\mathrm{max}$, but the qualitative behavior remains the same. We recover an analog behavior when comparing mass-gain and single-star models at identical values of $\Delta\Pi_1$ (see Fig.~\ref{fig:field_ratio_evo}). This difference in sensitivity is a direct consequence of the conservation of the $N$ profile during the mass-gain event (see Fig.~\ref{BruntVas}), which determines the global magnetic sensitivity parameter $\mathcal{I}$ (see Eq.~\ref{eq:struc_fac}). As a result, mass-gain models above the mass threshold show stronger magnetic signatures in their oscillations than single-star models at identical values for $B_\mathrm{avg}$. When estimating $B_\mathrm{avg}$ for a mass-gain candidate, adopting a mass-gain evolutionary history provides a tighter constraint of both the global magnetic sensitivity and $B_\mathrm{avg}$, as it allows for the simultaneous constraining of the core and envelope properties.

\section{Influence of strong magnetic fields}
\label{sec:strong_fields}

Moving away from the weak field limit, we now investigate the impact of strong magnetic fields on stellar oscillations under the assumption of a mass-gain history. Specifically, we assume core fields that reach or exceed the critical magnetic field strength $B_\mathrm{crit}$, the threshold above which the gravity-wave component of dipole mixed modes becomes converted into magnetic waves \citep[e.g.][]{Fuller15, Stello16a, Stello2016b, Lecoanet17, Loi20, Muller25}.

\subsection{Mode suppression due to magnetism}

To explain the observed population of RG stars with suppressed mixed dipole modes \citep[e.g.][]{Garcia14, Stello2016b}, the study of \cite{Fuller15} showed that magnetic fields can suppress mixed dipole modes if the magnetic field strength somewhere in the stellar interior exceeds a critical value of,
\begin{equation}
    B_{\mathrm{crit}}=\sqrt{\frac{\pi \rho}{2}}\frac{\left(2\pi\right)^2\nu^2 r}{N},
    \label{eq:b_crit}
\end{equation}
with \cite{Lecoanet17} further showing that this suppression happens due to the gravity wave component being converted into magnetic waves. In the case of RGs, the minimum critical field strength necessary to cause such a conversion,
\begin{equation}
    B_\mathrm{crit,min} = \min\left(B_\mathrm{crit}\left(r\right)\right).
\end{equation}
is located at the hydrogen-burning shell, where the Brunt-Väisälä frequency reaches its maximum (see Fig.~\ref{BruntVas}), owing to the $B_\mathrm{crit}\propto 1/N$ relation found in Eq.~\ref{eq:b_crit}.

\subsection{Dependence of the critical field strength $B_\mathrm{crit}$ on the Assumed Mass Gain History}
\label{sec:B_crit}

As the core structure is largely conserved during a mass-gain event \citep{Rui21}, we expect $B_\mathrm{crit,min}$ to differ between the single-star and mass-gain models. We focus in particular on higher model masses, where the global magnetic sensitivity between the two evolutionary scenarios diverges most strongly (see Sect~\ref{sec:magnetic_shifts} and Sect.~\ref{sec:stellar_structure}). Figure~\ref{fig:B_crit}, in which we assume $\nu=\nu_\mathrm{max}$ for all depicted models, confirms this expected behavior of $B_\mathrm{crit,min}$. While the single-star models exhibit the expected steep increase in $B_\mathrm{crit,min}$ for model masses above $1.5\,\mathrm{M_\odot}$ \citep[see][]{Fuller15, Stello2016b}, $B_\mathrm{crit,min}$ barely changes with model mass for the mass-gain models, owing to their similar core structure. 

Notably, $B_\mathrm{crit,min}$ remains in the few-hundred $\mathrm{kG}$ range for all mass-gain models with $M_\mathrm{init}<1.5\,M_\odot$, substantially below the $>1\,\mathrm{MG}$ values obtained for the corresponding high-mass single-star models. As shown by \cite{Stello2016b}, a significant fraction of \textit{Kepler} RG stars with masses $>1.6\,M_\odot$ (about $60\%$ in their sample) show depressed mixed dipole modes. To explain the observed mode suppression in these targets using the model of \cite{Fuller15}, very strong magnetic fields on the order of multiple hundred $\mathrm{kG}$ up to multiple $\mathrm{MG}$ would be required. Using the dipole mode visibility $V_1^2$ reported in the APOKASC3 catalog \citep{Stello2016b,Pinsonneault25} together with the single-star and mass-gain models, we recreate on the middle panel of Fig.~\ref{fig:depressed_mg_scenario} the Fig. 2 of \cite{Stello2016b}, in which they showed the parameter space occupied by $B_\mathrm{crit\min}$ for different values for $M$ and $\nu_\mathrm{max}$. We expand this analysis by calculating the parameter space of $B_\mathrm{crit,min}$ under the assumption of a mass-gain evolutionary history. For the mass-gain models, we assume a common initial mass of $M_\mathrm{init}=1.0\,M_\odot$. The resulting parameter space of $B_\mathrm{crit,min}$ for the mass-gain and single-star evolutionary history is presented in the left and right panels of Fig.~\ref{fig:depressed_mg_scenario}, respectively. We also show the ratio of $B_\mathrm{crit,min}$ between the single-star and mass-gain histories in the right panel of Fig.~\ref{fig:depressed_mg_scenario}. We find that adopting a $M_\mathrm{init}=1.0\,M_\odot$ mass-gain history reduces the critical field strength required to explain depressed dipole modes in RG stars with $M> 1.6\,M_\odot$ and $\nu_\mathrm{max} \gtrsim 170\,\mathrm{\mu Hz}$ from $B_\mathrm{crit,min} \gtrsim 1\, \mathrm{MG}$ to $B_\mathrm{crit,min}\lesssim 600\,\mathrm{kG}$. Albeit still being a considerably large value, this lies within the range of reported values for the weighted average of the radial field strength of RGB stars \citep[see][]{Deheuvels23}.

While not all depressed stars with masses exceeding $1.6\,M_\odot$ can necessarily be identified as mass-gain products, the fact that mass-gain evolutionary scenarios require substantially weaker magnetic fields to show suppression of mixed dipole modes suggests a promising explanation that warrants further exploration in future studies.

\section{Conclusion}
\label{sec:conclusion}

In this work, we quantified the degeneracy introduced to the local and global sensitivity of stellar oscillation modes to an internal magnetic field when different stellar evolutionary histories are considered for low- to intermediate-mass red giant branch solar-like oscillators. In our study, we compared stellar models that experienced a mass-gain event during the early RGB phase to stellar models showing a typical single-star evolution in a mass range of $1.1\,M_\odot$ to $2\,M_\odot$. We perform our analysis in two regimes: the weak field regime, where fields induce a magnetic frequency shift to mixed dipole modes, and the strong field regime, where fields are strong enough to suppress mixed dipole modes. Our results demonstrate that the assumed stellar evolutionary history has a strong impact when inferring the parameters of internal magnetic fields through asteroseismology.

In the weak field regime, we demonstrate that for a g-dominated mode with a given frequency $\nu$ and magnetic frequency shift $\delta\nu_\mathrm{mag}$, the inferred value of the weighted average of the radial magnetic field strength $\langle B_r^2\rangle^{1/2}$ obtained using the formulation for asymptotic gravity modes \citep{Bugnet21, Li22} strongly depends on the assumed evolutionary history for models of identical total mass. This effect originates from differences in the global magnetic sensitivity $\mathcal{I}$, which is closely linked to the Brunt-Väisälä frequency $N$ and the model density $\rho$ (see Eq.~\ref{eq:struc_fac}). These differences in $\mathcal{I}$ arise from the conservation of the core structure and thus the $N$ and $\rho$ profile of mass-gain models during the mass-gain event \citep{Rui21,Deheuvels_mg22}. We also find, that the profile of the asymptotic magnetic sensitivity kernel $K$ is conserved during a mass-gain event, owing to its dependence on $N$ and $\rho$ (see Eq.~\ref{eq:magnetic_kernel}).

We constrained the stellar core properties by evaluating $B_\mathrm{avg}$ for mass-gain models with an initial mass of $1.0\,M_\odot$ and single-star models of corresponding total mass in the range $1.1\,M_\odot$ to $2.0\,M_\odot$ and calculated the magnetic field ratio $R_\mathrm{B}$. We show that for identical $\Delta\Pi_1$, the field strength $\langle B_r^2\rangle^{1/2}$ obtained for the mass-gain and single-star models differs by up to $25\%$ when constraints on the model envelope properties are relaxed.

We constructed single-star and mass-gain model grids to estimate $\langle B_r^2\rangle^{1/2}$ for KIC 4350501 and KIC 9508757, two strong merger candidates identified in the literature \citep{Hatt24, Villate26}. We found acceptable mass-gain models with a fitting $\Delta\Pi_1/\Delta\nu/\nu_\mathrm{max}$ combination for both targets. For both KIC 4350501 and KIC 9508757, adopting a mass-gain evolutionary scenario reduces the range of possible values for $\langle B_r^2\rangle^{1/2}$ relative to the corresponding models of our single-star grid with fitting values for $\Delta\Pi_1$.

In the strong-field regime, we find that a mass-gain evolutionary history substantially reduces the minimum critical field strength $B_\mathrm{crit,min}$ required for dipole-mode suppression (up to $\approx80\%$ in the most extreme case). For stars with $M>1.6\,M_\odot$ and $\nu_\mathrm{max}\gtrsim 170\,\mathrm{\mu Hz}$, we find that $B_\mathrm{crit,min}$ decreases to values well below $1\,\mathrm{MG}$, which is significantly less than the values predicted by single-star models occupying the same region in the $\nu_\mathrm{max}-M$ parameter space. These results suggest that stars with $M>1.6\,M_\odot$ and high values for $\nu_\mathrm{max}$ that show depressed dipole mixed modes may represent promising mass-gain candidates. We thus propose that the investigation of the evolutionary history of these stars could prove interesting to properly constrain $B_\mathrm{crit,min}$.

We conclude that magnetic detections in both the weak and the strong field regimes can be systematically biased if the evolutionary history of the star is poorly constrained. We thus suggest the inclusion of stellar models that follow a non-single-star evolution in future studies that involve the probing of probable merger/mass-gain candidates that show magnetic signatures in their oscillations.

%%%%%%%%%%%%%%%%%%%%%%%%%%%%%%%%%%%%%%%%%%%%%%%%%%%%%%%%%%%%%%
\begin{acknowledgements}
   L. Buchele, L. Bugnet, and L. Einramhof gratefully acknowledge support from the European Research Council (ERC) under the Horizon Europe programme (Calcifer; PI Bugnet; Starting Grant agreement N$^\circ$101165631). While partially funded by the European Union, views and opinions expressed are, however, those of the authors only and do not necessarily reflect those of the European Union or the European Research Council. Neither the European Union nor the granting authority can be held responsible for them. L. Barrault acknowledges the support of the Austrian Academy of Sciences through the Doctoral Fellowship Programme (DOC) of the Austrian Academy of Sciences 27648. 
    P. G. Beck acknowledges support by the Spanish Ministry of Science, Innovation and Universities (MCIN) with the \textit{Ram{\'o}n\,y\,Cajal} fellowship (RYC-2021-033137-I). P. G. Beck acknowledge support from MCIN for the project \textit{PLAtoSOnG} from its grant PID2023-146453NB-I00. P. G. Beck was supported by the International Space Science Institute (ISSI) in Bern with the project\,24-629. PGB acknowledges financial support from the Agencia Estatal de Investigación (AEI) through the Severo Ochoa Centre of Excellence accreditation awarded to the Instituto de Astrofísica de Canarias, grant CEX2025-001609-S, funded by MICIU/AEI/10.13039/501100011033.
\end{acknowledgements}

\bibliographystyle{aa}
\bibliography{bibliography}

\begin{appendix}

\section{Additional model information}

\subsection{Employed model physics of our MESA setup}
\label{app:mesa_details}

In this Appendix, we intend to provide further information about the base setup we have employed in all MESA models used in this work. As already stated in Sect.~\ref{sec:model_info}, we base our MESA inlists on the work conducted by \cite{Rui21}.

We make use of the standard equation of state combination consisting of the OPAL/SCVH blend \citep{Rogers02}, HELM \citep{Timmes00}, FreeEOS \citep{Irwin12}, and SKYE \citep{Jermyn21}. For our opacity tables in the high and low temperature regime, we choose the standard solar abundances provided by \cite{Grevesse98} additionally supplementing them using the abundances reported by \cite{Ferguson05} in the low temperature regime.  

For our nuclear network, we employ the MESA standard \textsc{basic.net} network, which makes use of the nuclear reaction rates taken from JINA REACLIB \cite{Cyburt10} employing additional tabulated weak reaction rates \citep{Fuller85, Oda94, Langanke00}. 

For our model atmosphere, we are making use of a varying Eddington-Grey atmosphere description \citep{Eddington26}. To model mass loss due to stellar winds, we employ the Reimers mass loss scheme \citep{Reimers75}. We define the borders of the models convective and radiative zone via the Schwarzschild criterion \cite{Paxton13} further enabling predictive mixing. We use the mixing length description provided by \cite{Henyey65} and employ an exponential overshoot description \citep{Herwig00} with an uniform overshoot parameter $f_0=0.005$ across all models. We calculate our models as non rotating stars.

According to the recommendation stated by \cite{Buchele25}, we assume a value of $10^{-2}\,\mathrm{cm^2/s}$ for \textsc{overshoot\_D\_min}. We furthermore make use of the improved routine presented by \cite{Buchele26} to calculate $\Delta\Pi_1$.

\subsection{Additional figures}
\label{app:additional_figures}
\begin{figure*}[t]
    \centering
    \includegraphics[scale=0.83]{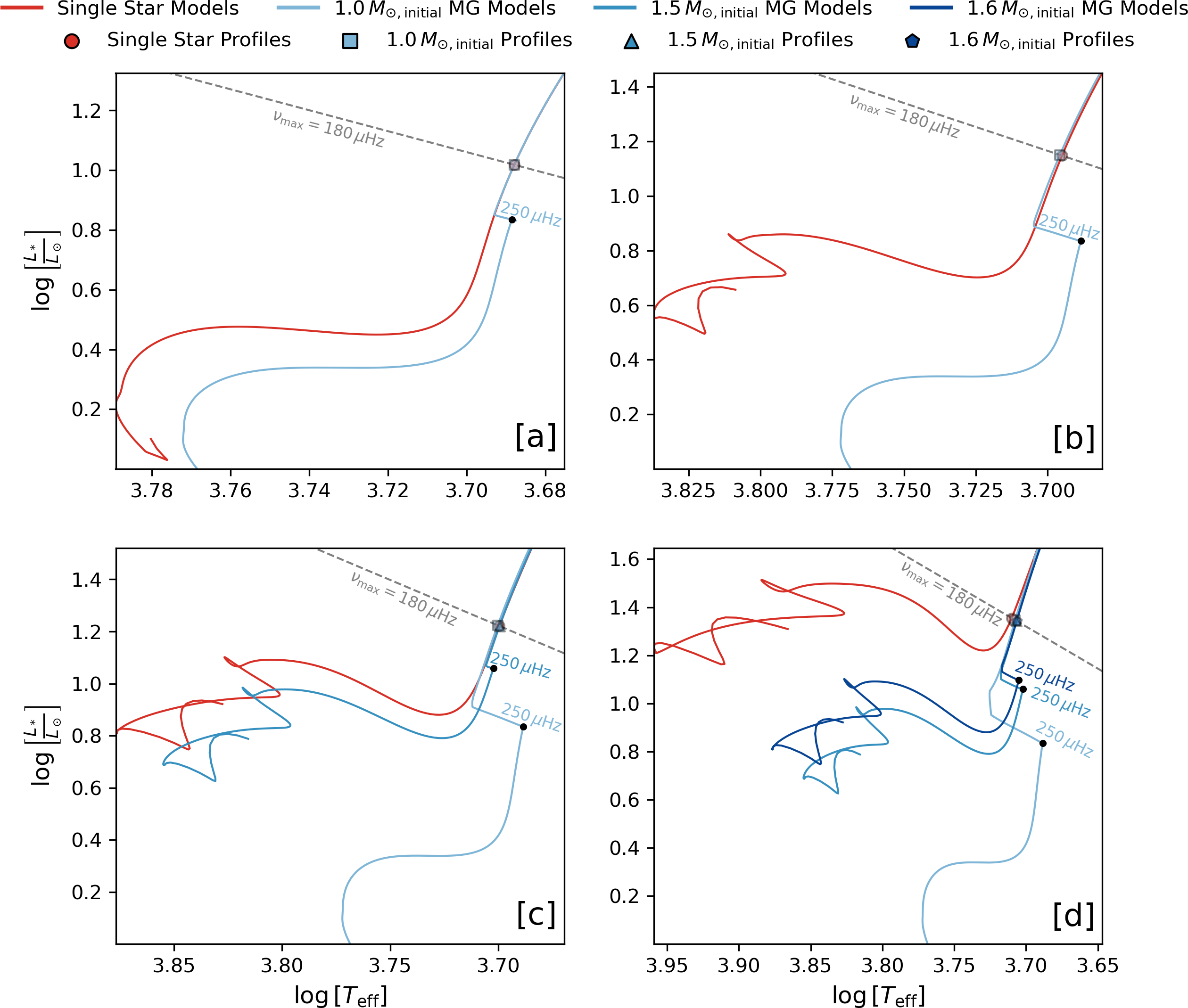}
    {\caption{Evolutionary tracks for the representative single-star and mass-gain models, with the points in the HR diagram corresponding to the onset of the mass-gain event marked by model-specific symbols. The stellar models depicted in plot $\left[a\right]$, $\left[b\right]$, $\left[c\right]$, and $\left[d\right]$ possess a total (post-gain) mass of $1.1\,M_\odot$, $1.5\, M_\odot$, $1.6\, M_\odot$ and $2\,M_\odot$ respectively. The global power excess $\nu_\mathrm{max}$ is marked at a value of $180\, \mathrm{\mu Hz}$ for all model tracks, making use of a rewritten form of the scaling relation introduced by \cite{Kjeldsen95}. The mass used to plot the tracks for $\mathrm{\nu_\mathrm{max}}$ corresponds to the total (post-gain) mass of the models depicted in the respective panel.}\label{FigHRD}}
\end{figure*}

\begin{figure*}[t]
\centering
\begin{minipage}[c]{0.73\textwidth}
    \centering
    \includegraphics[width=\linewidth]{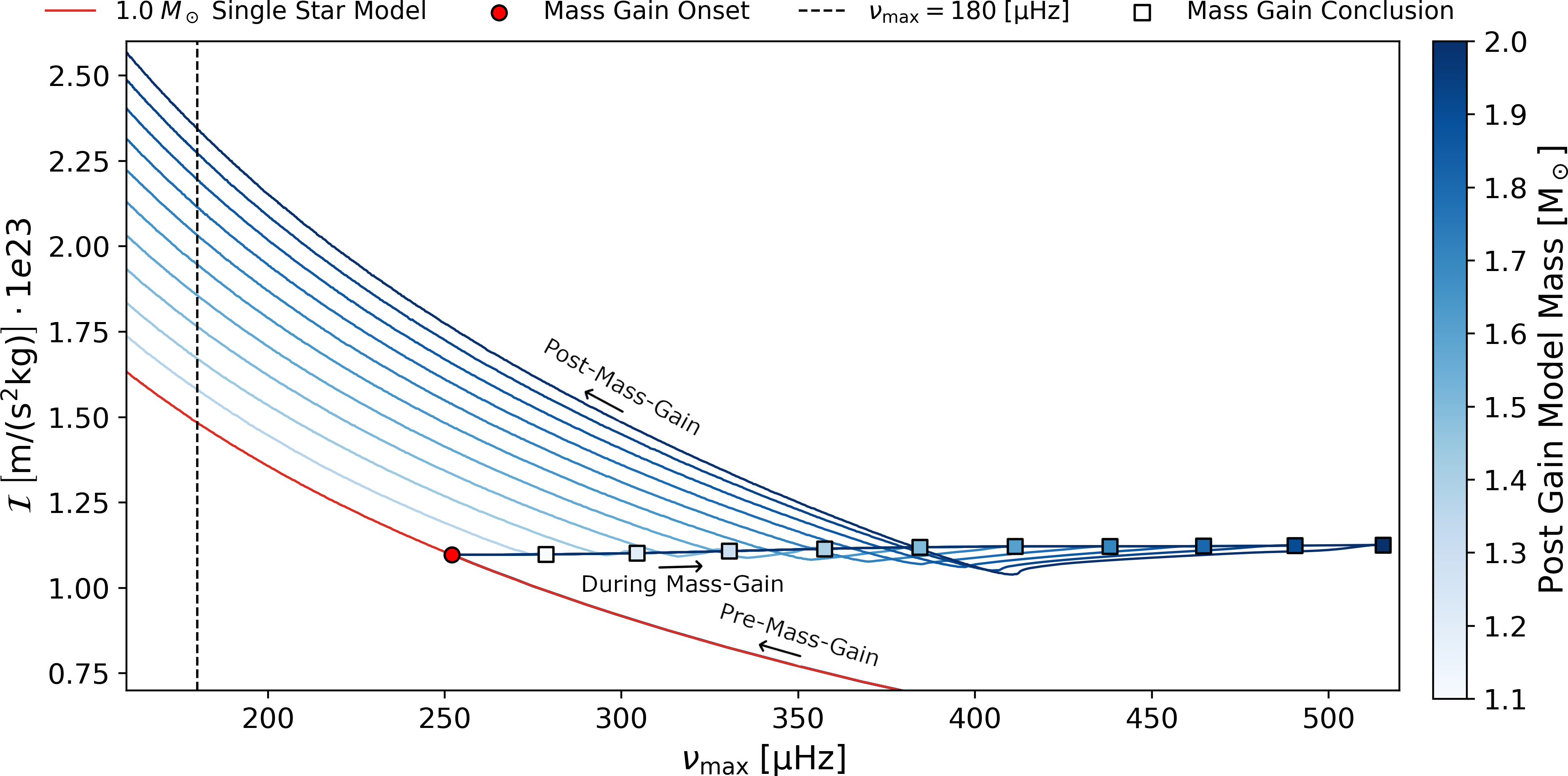}
\end{minipage}\hfill
\begin{minipage}[c]{0.25\textwidth}
    \captionsetup{type=figure, labelfont=bf, font=small,
                  justification=raggedright, singlelinecheck=false}
    \captionof{figure}{Relation between $\nu_\mathrm{max}$ and $\mathcal{I}$ for mass-gain models of initial mass $1.0\;M_\odot$ and post gain masses of $M_\mathrm{total}\in\left[1.1\;M_\odot,2.0\;M_\odot\right]$ (blue model tracks) and a $1.0\;M_\odot$ single-star model. For the ease of the reader, the direction of evolution is marked by arrows that are labeled according to the evolutionary stage of the mass-gain models}
    \label{fig:numax_vs_sf}
\end{minipage}
\end{figure*}

\begin{figure*}[t]
    \centering
    \includegraphics[scale=0.6]{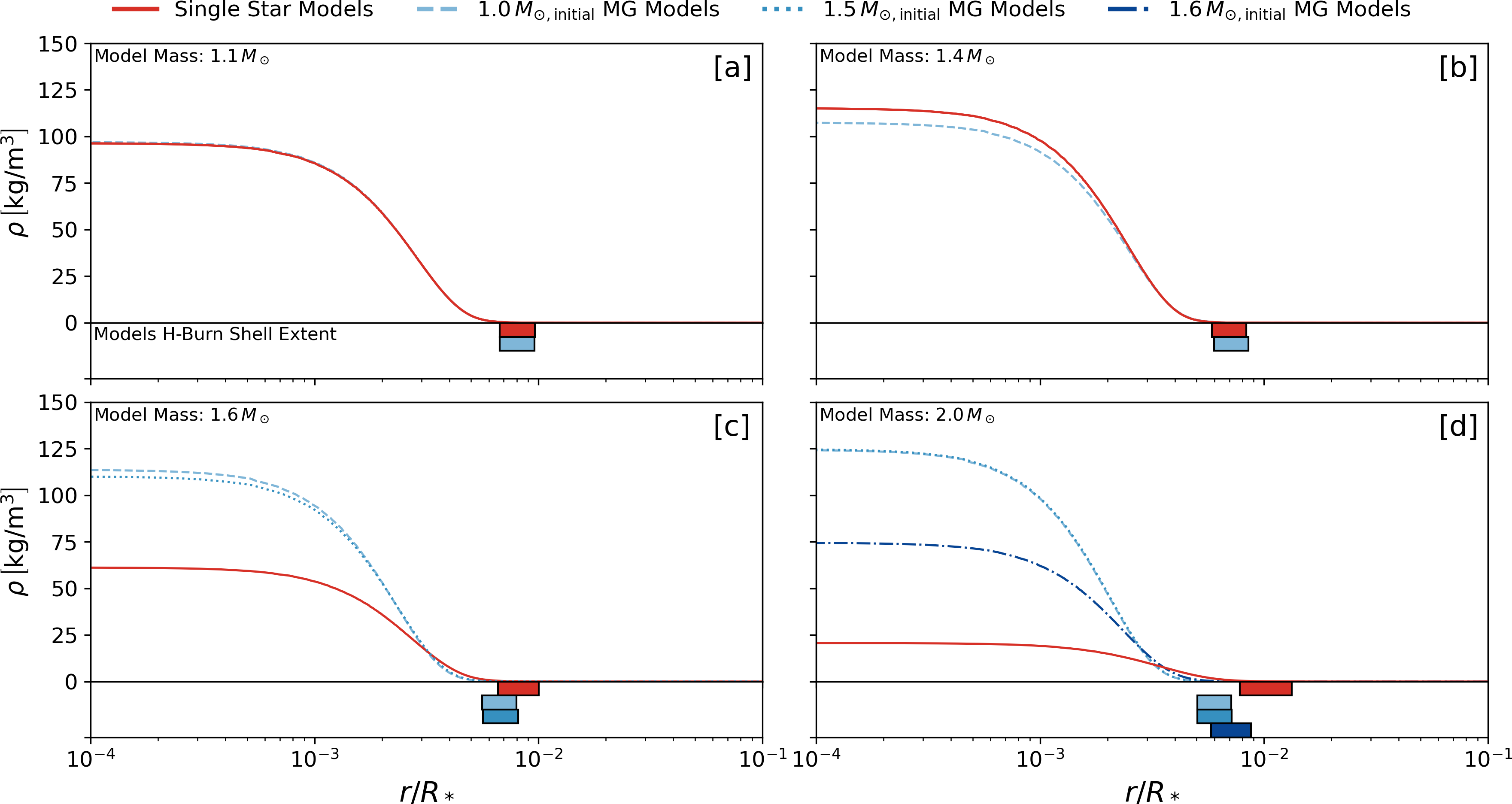}
    {\caption{The model density $\rho$ as well as the model-specific extent of the hydrogen burning shell obtained for the representative single-star and mass-gain models at a value for the frequency of the global power excess of $\nu_\mathrm{max}=180\,\mathrm{\mu Hz}$. The colored boxes at the bottom of each panel signify the extent of the hydrogen-burning shell of the respective single-star or mass-gain model. The labels $\left[a\right]$, $\left[b\right]$, $\left[c\right]$, and $\left[d\right]$ correspond to the model labeling introduced in Tab.~\ref{tab:modelproperties}.}\label{fig:internal_density}}
\end{figure*}

\begin{figure*}[t]
    \centering
    \includegraphics[scale=0.6]{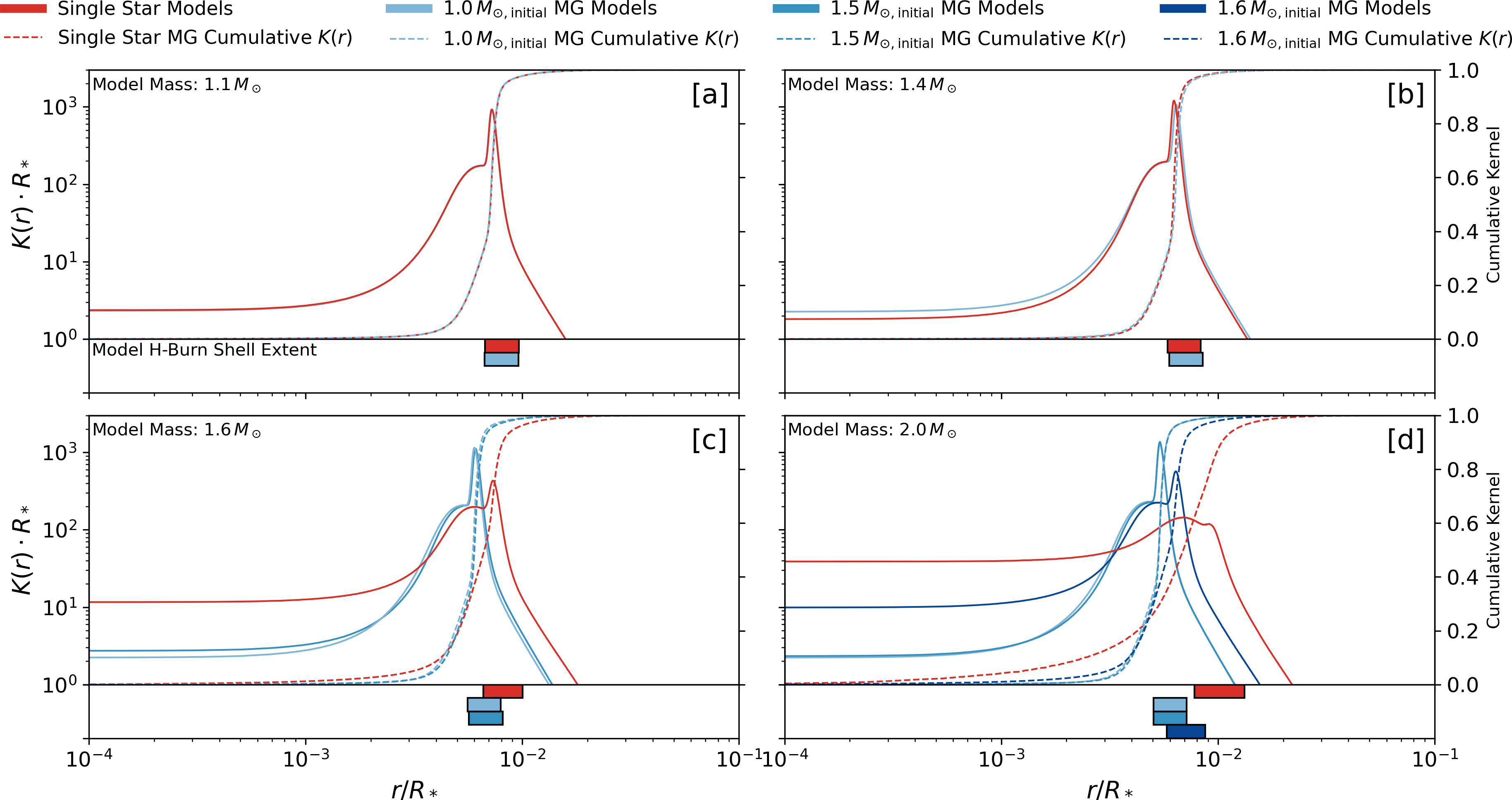}
    {\caption{The magnetic kernel function $K$ as well as the cumulative magnetic kernel function and the model specific extent of the hydrogen burning shell obtained for the representative single-star and mass-gain models. The labels $\left[a\right]$, $\left[b\right]$, $\left[c\right]$, and $\left[d\right]$ correspond to the model labeling introduced in Tab.~\ref{tab:modelproperties}.}\label{fig:mag_kern}}
\end{figure*}

In this Appendix, we present additional figures that complement the findings in our work. We provide a Hertzsprung-Russell diagram in Fig.~\ref{FigHRD} showing the MESA model tracks of the representative stellar models listed in Tab.~\ref{tab:modelproperties}. We also provide a diagram showing the $\nu_\mathrm{max}-\mathcal{I}$ relation for the $M_\mathrm{init}=1.0\,M_\odot$ mass-gain models in Fig.~\ref{fig:numax_vs_sf}, and the model density and magnetic sensitivity kernel profiles for the representative models listed in Tab.~\ref{tab:modelproperties} in Fig.~\ref{fig:internal_density} and Fig.~\ref{fig:mag_kern} respectively. We furthermore present the stellar radii and effective temperatures of the single-star and mass-gain models used in Sect.~\ref{sec:magnetic_shifts} and Sect.~\ref{sec:stellar_structure} in Tab.~\ref{tab:stellar_model_params}.

\begin{table*}[t]
\centering
\caption{Effective temperatures and stellar radii of our stellar models.}
\label{tab:stellar_model_params}
\begingroup
\begin{tabular}{c|cc|cc|cc|cc}
\hline\hline
$M_\mathrm{total}$ & \multicolumn{2}{c|}{Single} & \multicolumn{2}{c|}{Mass-Gain ($M_\mathrm{init}=1\,M_\odot$)} & \multicolumn{2}{c|}{Mass-Gain ($M_\mathrm{init}=1.5\,M_\odot$)} & \multicolumn{2}{c}{Mass-Gain ($M_\mathrm{init}=1.6\,M_\odot$)} \\
$[M_\odot]$ & $T_{\rm eff}$ [K] & $R$ $[R_\odot]$ & $T_{\rm eff}$ [K] & $R$ $[R_\odot]$ & $T_{\rm eff}$ [K] & $R$ $[R_\odot]$ & $T_{\rm eff}$ [K] & $R$ $[R_\odot]$ \\
\hline
1.0 & 4839 & 4.327 & / & / & / & / & / & / \\
1.1 & 4874 & 4.530 & 4873 & 4.535 & / & / & / & / \\
1.2 & 4900 & 4.724 & 4905 & 4.728 & / & / & / & / \\
1.3 & 4928 & 4.914 & 4935 & 4.912 & / & / & / & / \\
1.4 & 4954 & 5.093 & 4963 & 5.093 & / & / & / & / \\
1.5 & 4980 & 5.263 & 4989 & 5.267 & / & / & / & / \\
1.6 & 5006 & 5.428 & 5014 & 5.435 & 5004 & 5.435 & / & / \\
1.7 & 5034 & 5.598 & 5037 & 5.592 & 5027 & 5.595 & 5027 & 5.595 \\
1.8 & 5065 & 5.739 & 5059 & 5.749 & 5048 & 5.748 & 5048 & 5.754 \\
1.9 & 5097 & 5.876 & 5079 & 5.900 & 5069 & 5.904 & 5067 & 5.906 \\
2.0 & 5130 & 6.015 & 5099 & 6.048 & 5088 & 6.051 & 5086 & 6.056 \\
\hline
\end{tabular}
\tablefoot{The parameters provided correspond to the single star and mass-gain models used for our analysis in Sect.~\ref{sec:magnetic_shifts} and Sect.~\ref{sec:stellar_structure}. The mass value in the $M_\mathrm{total}$ columns corresponds to the model mass in the case of the single-star models, and to the model mass after the mass-gain event in the case of the mass-gain models. All model parameters listed are taken at $\nu_\mathrm{max}=180\,\mathrm{\mu Hz}$.}
\endgroup
\end{table*}

\end{appendix}
\end{document}